\documentclass[
aps,
prd,
superscriptaddress,
nofootinbib,
floatfix]
{revtex4}
\usepackage{graphicx,
longtable}
\begin{document}
\title{Neutrino mass hierarchy and precision physics with medium-baseline reactors:\\ Impact of energy-scale and flux-shape uncertainties}

%
\author{        F.~Capozzi}
\affiliation{   Dipartimento Interateneo di Fisica ``Michelangelo Merlin,'' 
               Via Amendola 173, 70126 Bari, Italy}%
\affiliation{   Istituto Nazionale di Fisica Nucleare, Sezione di Bari, 
               Via Orabona 4, 70126 Bari, Italy}
\author{        E.~Lisi}
\affiliation{   Istituto Nazionale di Fisica Nucleare, Sezione di Bari, 
               Via Orabona 4, 70126 Bari, Italy}
\author{        A.~Marrone}
\affiliation{   Dipartimento Interateneo di Fisica ``Michelangelo Merlin,'' 
               Via Amendola 173, 70126 Bari, Italy}%
\affiliation{   Istituto Nazionale di Fisica Nucleare, Sezione di Bari, 
               Via Orabona 4, 70126 Bari, Italy}
\begin{abstract}
\vspace*{1cm}
Nuclear reactors provide intense sources of electron antineutrinos, characterized by few-MeV energy $E$ and unoscillated spectral shape $\Phi(E)$. High-statistics observations of reactor neutrino oscillations over medium-baseline distances $L\sim O(50)$~km would provide unprecedented opportunities to probe both the 
long-wavelength mass-mixing parameters ($\delta m^2$ and $\theta_{12}$) and the short-wavelength ones ($\Delta m^2_{ee}$ and $\theta_{13}$), together 
with the subtle interference effects associated with the neutrino mass hierarchy (either normal  or inverted). In a given experimental 
setting---here taken as in the JUNO project for definiteness---the achievable hierarchy sensitivity and parameter accuracy depend not only on the accumulated statistics but
also on systematic uncertainties, which include (but are not limited to) the mass-mixing priors and the normalizations of signals and backgrounds. We examine, in addition, the effect of introducing smooth deformations of the detector energy scale, $E\to E'(E)$, and of the reactor flux shape, $\Phi(E)\to \Phi'(E)$, within reasonable error bands inspired by  state-of-the-art estimates. It turns out that  energy-scale and flux-shape systematics can noticeably affect the performance of a JUNO-like experiment, both on the hierarchy discrimination and on precision oscillation physics.
It is shown that a significant reduction of the assumed 
energy-scale and flux-shape uncertainties  (by, say, a factor of 2) would be highly beneficial to the physics program of medium-baseline reactor projects. Our results also shed some light on the role of the inverse-beta decay threshold, of geoneutrino backgrounds, and of matter effects in the analysis of future reactor oscillation data.   
\end{abstract}
\medskip
\pacs{
14.60.Pq,
13.15.+g,
28.50.Hw} 
\maketitle

\section{Introduction}

The currently established neutrino oscillation phenomenology can be interpreted in a
three-generation framework, whose relevant parameters are 
three mixing angles $(\theta_{12},\,\theta_{13},\theta_{23})$, one possible CP-violating phase $\delta$, and two neutrino squared mass differences 
$(\delta m^2,\,\pm \Delta m^2)$, where the latter sign distinguishes the cases of normal hierarchy (NH,~$+$) and inverted 
hierarchy (IH,~$-$) for the neutrino mass spectrum \cite{PDGR}. In reactor neutrino experiments, for a given neutrino energy $E$ and baseline $L$,
the $\overline\nu_e$ disappearance probability depends only on a subset of parameters, 
\begin{equation}
\label{Pee}
P(\overline\nu_e\to\overline\nu_e) = P_{ee}(\theta_{13},\,\theta_{12},\,\delta m^2,\,\pm\Delta m^2)\ ,
\end{equation}
and is generally not
invariant under a  swap of the $\Delta m^2$ sign, thus providing some sensitivity to the mass hierarchy as noted in \cite{Fo01}. 
Hierarchy effects vanish in the limits of short baselines [$L\lesssim {\cal O}(1)$~km] and of long baselines   
[$L\gtrsim {\cal O}(10^2)$~km], where the $P_{ee}$ arguments reduce to ($\theta_{13},\,|\Delta m^2|$) and 
$(\theta_{13},\,\theta_{12},\,\delta m^2)$, respectively. However, at medium baselines $L$ (few tens of km), the 
hierarchy sensitivity can be recovered in high-statistics reactor experiments, provided that interference effects
between short- and long-wavelength oscillations are resolved, as  originally proposed in \cite{Pe02}. 

Such a possibility is being seriously investigated experimentally,
as one of the main physics goals of the Jiangmen Underground Neutrino Observatory in China \cite{JUN1} (JUNO, in construction), 
and of the project RENO-50 proposed in Korea \cite{RE50}. 
A number of papers have examined the stringent conditions needed to discriminate the hierarchy in this class of reactor experiments, including high energy resolution, small dispersion of multiple core baselines, good control of energy-scale nonlinearities, and reduction of systematics related to oscillation parameters and normalizations. Satisfying these conditions would also provide more precise measurements of
the parameters ($\theta_{12},\,\delta m^2)$ \cite{Gosw,Min1} and of $\Delta m^2$. We refer the reader to \cite{Vo15,Vo16} for recent reviews of this  field
of research and of the related bibliography. 

In this context, it has been realized that the shape of the (unoscillated) reactor neutrino flux $\Phi(E)$ may not be known with the desirable
accuracy, as recently demonstrated by the unexpected spectral features consistently found in the current
short-baseline experiments RENO \cite{BRen}, Double Chooz \cite{BDou} and Daya Bay 
\cite{BDay}. In particular, an event excess (sometimes dubbed as a ``bump'' or ``shoulder'' in the spectrum) clearly emerges around $E\sim 5$--7~MeV, with respect to 
widely adopted theoretical predictions \cite{Muel,Hube}. 
Very recent calculations provide possible (although still partial) interpretations of the subtle nuclear effects which may be responsible for these 
(and possibly other) unexpected spectral features  \cite{Dwye,Haye,HuGe}, but a thorough understanding 
of their origin has not yet been achieved. 

In current short-baseline experiments, endowed with near and far detectors, 
poorly understood spectral features largely cancel in near-to-far flux ratios, and do not significantly affect the 
measurement of the dominant parameters $(\theta_{13},|\Delta m^2|)$ \cite{DB15}. However, in the absence 
of a near detector, a medium-baseline experiment such as JUNO must rely on absolute estimates of the flux spectrum $\Phi(E)$
and of its associated uncertainties, which may affect both the sensitivity to the hierarchy and the 
accuracy of the ($\theta_{12},\,\delta m^2,\,|\Delta m^2|$) measurements. 

A related and subtle issue concerns energy-scale variations 
$E\to E'(E)$ \cite{Park} which, in principle, may be nonlinearly engineered to produce a hierarchy
ambiguity  \cite{Qian}. Although the energy-scale issue can be kept under control by calibration constraints at subpercent level \cite{Unam}, 
specific combinations of energy variations $E\to E'(E)$  and spectral deviations $\Phi(E)\to\Phi'(E)$  
might represent a more subtle threat to the hierarchy discrimination \cite{Ours}. In general, delicate statistical aspects---related to the  
treatment of admissible spectral deformations---are now emerging in neutrino oscillation searches, mirroring the evolution
of other fields of physics from the discovery phase to the precision era, as remarked in \cite{PING}.

Within this scenario, and building upon our previous work \cite{Ours}, we present herein a systematic analysis of the combined effects 
of energy-scale nonlinearities $E\to E'(E)$ and flux-shape uncertainties $\Phi(E)\to\Phi'(E)$, assuming a reference JUNO-like medium-baseline
reactor neutrino setting.
 The structure of our paper is as follows: In Sec.~II  we describe the adopted notation and methodology. In Sec.~III we discuss the effects
of energy-scale and flux-shape  uncertainties on the hierarchy discrimination, while in Sec.~IV we show their impact on precision 
measurements of the oscillation parameters. In Sec.~V we repeat the analysis in a prospective scenario where 
the energy-scale and flux-shape uncertainties are reduced by a factor of two. 
We also comment on the role of the inverse-beta decay threshold, geoneutrino backgrounds, and matter effects.   
A brief summary of the results, and perspectives for further work, is presented in Sec.~VI.

\section{Notation and Methodology}

We generally adopt the same  notation and inputs as in \cite{Ours} (to which the reader is referred for details), 
and comment below only about those aspects that are new, modified, or relevant for the present analysis. In particular,
we discuss the statistical approach used to characterize and include energy-scale and flux-shape uncertainties. 
We remark that the adopted experimental setup \cite{Ours}, as taken from \cite{Unam},
is basically the same as reported in a recent publication from the JUNO Collaboration  \cite{JUN1}.

\subsection{Neutrino oscillation parameters and priors}

Neutrino oscillation probabilities can be expressed in terms of neutrino squared mass differences ($\Delta m^2_{ji}=m^2_j-m^2_i$) and
trigonometric functions of the mixing angles $\theta_{ij}$ (e.g., $s_{ij}=\sin\theta_{ij}$,  $c_{ij}=\cos\theta_{ij}$). In medium-baseline reactor experiments, there are four relevant parameters: $s^2_{12}$, $s^2_{13}$, $\delta m^2=\Delta m^2_{21}$, and $\Delta m^2_{ee}$  defined as \cite{Gouv,Nuno,Mina}
\begin{equation}
\Delta m^2_{ee} = \Delta m^2 \pm \frac{1}{2}(c^2_{12}-s^2_{12})\delta m^2\ , 
\label{Deltam2ee}
\end{equation}
where the upper (lower) sign refers to NH (IH), while $\Delta m^2$ is defined as \cite{Ours,GFit}
\begin{equation}
\Delta m^2 = \frac{1}{2}\left|\Delta m^2_{31}+\Delta m^2_{32}\right|>0\ .
\end{equation}
An accurate analytical expression for the oscillation probability $P_{ee}$, including effects due to propagation in constant-density matter and to multiple reactor cores, can be found in \cite{Ours} [see Eqs.~(58) and (59) therein].

We assume the following reasonable priors (central values and $\pm1\sigma$ errors) 
for the above parameters, at the start of a JUNO-like experiment:
\begin{eqnarray}
s^2_{12} &=& (3.08\pm  0.17)\times 10^{-1}\ , \\
\delta m^2 &=& (7.54\pm 0.20)\times 10^{-5}\  \mathrm{eV}^2\ , \\
s^2_{13} &=& (2.20 \pm 0.08)\times 10^{-2}\ , \\
\Delta m^2_{ee} &=& (2.40 \pm 0.05)\times 10^{-3}\    \mathrm{eV}^2\ .
\end{eqnarray}
The ($s^2_{12},\,\delta m^2$) priors---unlikely to change significantly in the near future---are taken from the global fit in \cite{GFit}, with errors defined as $1/6$ of the $\pm 3\sigma$ range. 
 The $s^2_{13}$ central value is a bit lower than in \cite{GFit}, as suggested by recent reactor results \cite{RE50,BDou,DB15}, and is also endowed with
a smaller $\pm 1\sigma$ error, representative of the final accuracy expected in Daya Bay \cite{Wang}. 
Finally, the $\Delta m^2_{ee}$ central value is also in ballpark of the current global fits \cite{GFit,Fit1,Fit2}, but with a somewhat smaller fractional error than in \cite{GFit}
($2.0\%$ instead of $2.6\%$) as it can be expected from near-future improvements in short-baseline reactor \cite{Wang} and long-baseline accelerator experiments   \cite{T2K1}.

\subsection{Reactor and geoneutrino spectra, energy resolution and thresholds}

Concerning the reactor neutrino fluxes (from both medium-baseline and far sources), we use the same average fuel component and overall normalization as
in \cite{Ours}, but we alter the energy profile to include the newly discovered spectral feature at $E\sim 5$--7~MeV  \cite{BRen,BDou,BDay}. In
particular, we multiply the unoscillated reactor spectrum in \cite{Ours} by a smoothed version of the bin-to-bin ratio (Daya Bay)/(Huber~+~Mueller) reported in \cite{Wang}, which effectively accounts for the spectral bump feature as observed in Daya Bay \cite{BDay}. 

The normalization of the U and Th geoneutrino background 
fluxes in JUNO is slightly increased with respect to \cite{Ours} by 7\% and 9\%, respectively, in order to match
the most recent estimates in \cite{JGeo}. We also update the associated errors, by conservatively taking the largest of the asymmetric $1\sigma$  uncertainties from \cite{JGeo}, namely, $\pm 20\%$  for the U flux and $\pm 27\%$ for the Th flux. 

Finally, with respect to  \cite{Ours}, we  slightly update the energy resolution width $\sigma_e$ as reported in \cite{Wang},
\begin{equation}
\frac{\sigma_e(E_e)}{E_e+m_e} = \frac{2.57\times 10^{-2}}{\sqrt{(E_e+m_e)/\mathrm{MeV}}} + 0.18\times 10^{-2}\ ,
\label{width}
\end{equation}
where $E_e+m_e$ is the total (true) visible energy of the inverse beta decay (IBD) event, as a sum of the total positron energy $E_e$ and electron mass $m_e$. 
We remind the reader of the following \cite{Ours}: 
(1) the finite resolution width smears the observed visible energy $E_\mathrm{vis}$ around its true value $E_e+m_e$; (2)
the neutrino energy $E$ has an IBD kinematical threshold $E\geq 1.806$~MeV; (3) the parent neutrino energy $E$ and the observed visible energy of the event $E_\mathrm{vis}$
are approximately related by
\begin{equation}
E\simeq E_\mathrm{vis}+0.78~\mathrm{MeV}\ ,
\label{shift}
\end{equation}
up to nucleon recoil and energy resolution effects (which we accurately include in the calculations of energy spectra \cite{Ours}); and (4) the visible energy threshold is thus $E_\mathrm{vis}\gtrsim 1$~MeV. 

\begin{figure}[b]
\begin{minipage}[c]{0.96\textwidth}
\includegraphics[width=0.38\textwidth]{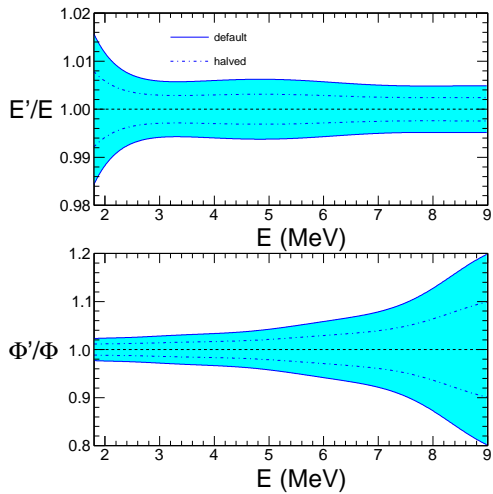}
\caption{\label{fig01}\footnotesize Default $\pm1\sigma$ error bands assumed for energy-scale deviations $E'/E$ (top panel) and flux-shape variations $\Phi'/\Phi$ (bottom panel), in terms of the neutrino energy $E$. Bands with halved errors are also shown (dot-dashed lines in both panels).}
\end{minipage}
\end{figure}

\subsection{Error bands for energy-scale and flux-shape deformations}

In first approximation, we assume a JUNO energy-scale uncertainty comparable to the Daya Bay one. For current Daya Bay data, 
the $1\sigma$ error band of admissible deviations in the reconstructed/true visible energy ratio has been shown in \cite{DB15} and (with slightly smaller width) in 
\cite{Err1,Err2}.  We have translated the bands in \cite{Err1,Err2} into relative deviations $E'/E$ 
for the neutrino energy via Eq.~(\ref{shift}).
Asymmetric $1\sigma$ uncertainties have been symmetrized to the largest between $+1\sigma$ and $-1\sigma$. Figure~1 (top panel) shows, 
in color, the resulting 
energy-scale error band (at $\pm 1\sigma$ in $E'/E$), as a function  of the parent neutrino energy $E$. Besides this ``default'' band, we shall 
also consider a more optimistic case with ``halved'' errors (dot-dashed lines in the top panel of Fig.~1), in view of dedicated energy
calibration campaigns expected in JUNO. 

Concerning the flux-shape uncertainties of the unoscillated reactor spectrum $\Phi(E)$, we assume that $\Phi'(E)/\Phi(E)$ deviations
are constrained by  the $\pm 1\sigma$ error bands estimated in \cite{Dwye}. We have smoothed out and symmetrized the  bands in \cite{Dwye}, as reported in
Fig.~1 (bottom panel) in terms of the neutrino energy $E$. Since the issue of reactor spectral shapes is still highly debated, the 
$\Phi'/\Phi$ error band should be taken as merely indicative of the current level of theoretical uncertainties.
The high statistics accumulated in the present generation of short-baseline reactor experiments will certainly help to
constrain any model of reactor spectra and, indeed, the current size of systematic shape uncertainties estimated in Daya Bay \cite{DB15} already seems to
be a factor of 2 smaller than in \cite{Dwye}, although a detailed assessment has not yet been published. For this reason,
also in the analysis for flux-shape uncertainties, we shall consider the more optimistic case of halved theoretical errors (dot-dashed lines in the bottom panel of Fig.~1).

A final remark is in order. In the absence of a detailed characterization of the error bands in Fig.~1, we simply
assume that they  scale linearly with $n\sigma$.
We also neglect, for lack of published
information, possible error correlations at different energies. Although some correlations are known to exist, 
as a result of underlying models for both the energy scale nonlinearities \cite{DB15,Err1,Err2} and the reactor spectra \cite{Muel,Hube,Dwye,Haye}, their 
impact should not be overemphasized at this stage. Indeed, the recently observed, localized ``bump'' feature largely exceeds the 
estimated errors and covariances that where thought to characterize the spectra a few years ago \cite{Muel,Hube}. In this sense, 
neglecting possible covariances in Fig.~1 should lead to conservative results. A more refined analysis will be possible when 
such error bands will be determined more precisely, and endowed with point-to-point correlation functions.

\subsection{Statistical approach}

As in \cite{Ours}, we calculate the ``true'' event spectrum $S^*(E_\mathrm{vis})$ by assuming the central values of 
the oscillation parameters reported 
in Sec.~II~A, for either NH or IH. Such a spectrum $S^*$ represents the ``experimental data'', to be
compared with a family of spectra $S(E_\mathrm{vis})$,
obtained by varying the continuous parameters ($\delta m^2,\,\Delta m^2_{ee},\,\theta_{12},\,\theta_{13}$), in either the same 
or the opposite hierarchy. The comparison is performed   
in terms of a $\chi^2$ function that contains statistical, parametric, and systematic components (see also \cite{Ge12}),
\begin{equation}
\chi^2 = \chi^2_\mathrm{stat} + \chi^2_\mathrm{par} + \chi^2_\mathrm{sys}\ .
\end{equation}
The $\chi^2_\mathrm{stat}$ term (which embeds statistical fluctuations) is the same as in \cite{Ours}. The $\chi^2_\mathrm{par}$ term (which 
embeds penalties for the oscillation parameters) is also unchanged, apart from the numerical priors, here taken from Sec.~II~A above. 

The $\chi^2_\mathrm{sys}$ term contains, as in \cite{Ours},  
two penalties related to the geoneutrino flux normalizations,
\begin{equation}
\chi^2_\mathrm{geo} = \sum_{j=\mathrm{U,Th}} \left(\frac{f_j-1}{s_j}\right)^2 \in \chi^2_\mathrm{sys}\ ,
\label{chisys}
\end{equation}
where we take $s_\mathrm{U}=0.20$ and $s_\mathrm{Th}=0.27$, as described in Sec.~II~B. 
However, while in \cite{Ours} the $\chi^2_\mathrm{sys}$ term was completed by just another penalty 
for the overall reactor flux normalization, now it must be supplemented by appropriate penalties for
energy-scale and flux-shape deformations.

To this purpose, we consider smooth deformations of the energy scale $E\to E'(E)$
(that we assume to act upon the ``experimental spectrum'' $S^*$) and of the flux shape $\Phi(E)\to\Phi'(E)$
(that we assume to act upon the ``theoretical spectrum'' $S$), in terms of generic polynomials in $E$ (in MeV),
\begin{eqnarray}
\frac{E'}{E} &=& 1+\sum_{i=0}^{k} \alpha_i E^i = 1+\delta_E(E)\ , \\
\frac{\Phi'(E)}{\Phi(E)} &=& 1+\sum_{j=0}^{h} \beta_j E^j= 1+\delta_\Phi(E)\ , 
\end{eqnarray}
with $h$ and $k$ increasing until stable results are reached \cite{Expl}. Note that the trivial
cases $h=0$ and $k=0$ correspond, respectively, to an overall renormalization 
of the energy scale [$E'=(1+\alpha_0) E$] and of the reactor spectrum [$\Phi'=(1+\beta_0) \Phi$].

With reference to Fig.~1, let us denote the boundaries of the $1\sigma$ error bands in Fig.~1 as $1\pm S_E(E)$ for the upper panel, and 
as $1\pm S_\Phi(E)$ 
for the lower panel. Then we define two new systematic penalties, in terms of the largest relative deviation associated with each polynomial:
\begin{eqnarray}
\chi^2_E &=& \max_E \left|\frac{\delta_E(E)}{S_E(E)}\right|^2 , \\
\chi^2_\Phi &=& \max_E \left|\frac{\delta_\Phi(E)}{S_\Phi(E)}\right|^2 . 
\end{eqnarray}
In other words, if the polynomial function $\delta_E(E)$ ``touches'' the $n\sigma$ error band boundary $ n\times S_E(E)$, 
its contribution to the $\chi^2_\mathrm{sys}$ term is assumed  to be $n^2$, and similarly for $\delta_\Phi(E)$ and $S_\Phi(E)$. 
Equivalently, the $\pm 1\sigma$ bands in Fig.~1 are assumed to be the envelope of all possible systematic deviations at the $1\sigma$ level,
and similarly for $n\sigma$.
Such a $\chi^2$ characterization of energy-scale and spectral-shape errors is both intuitive and conservative,
as appropriate to an exploratory analysis. As previously remarked, more refined definitions of $\chi^2_\mathrm{sys}$ will be possible in the future, in terms of energy-dependent cross correlations.

We have found that our results, to be discussed in the next sections, become numerically stable already for fifth-order polynomials,
which are taken as a default choice for all the following figures. 
Therefore, in general, the $\chi^2$ minimization requires
scanning a 18-dimensional parameter space, including four oscillation parameters $(s^2_{12},\,s^2_{13},\,\delta m^2,\,\Delta m^2_{ee})$,
 two geoneutrino flux normalizations $(f_\mathrm{U},\,f_\mathrm{Th})$, and twelve polynomial coefficients $(\alpha_0,\dots,\alpha_5)$ 
and $(\beta_0,\dots,\beta_5)$. [We have also cross-checked the numerical results by using different and independent minimization methods.] 

For the sake of the discussion,  we shall also consider cases with reduced dimensionality,
as obtained by setting to zero the coefficients $\alpha_i$
or $\beta_j$. However, in our analysis, the specific coefficient $\beta_0$ is never zeroed a priori, since it parametrizes
a floating normalization for the reactor flux, $\Phi\to \Phi(1+\beta_0)$. In particular,
we shall consider the following cases, in order of increasing number of free parameters: 
\begin{itemize}
\item oscillation + normalizations: $(s^2_{12},\,s^2_{13},\,\delta m^2,\,\Delta m^2_{ee})$ + $(f_\mathrm{U},\,f_\mathrm{Th})$ + $(\beta_0)$ = 7 parameters;
\item osc.\ + norm.\ + energy scale: as above + $(\alpha_0,\dots,\alpha_5)$ = 13 parameters; 
\item osc.\ + norm.\ + energy scale + flux shape: as above  + $(\beta_1,\dots,\beta_5)$ = 18 parameters.
\end{itemize}
In the first two cases, from the definition of $\chi^2_\Phi$, the $1\sigma$ error associated with $\beta_0$ 
coincides to the smallest error band width in Fig.~1 (bottom panel), i.e.\ to $\sim 2.3\%$, 
which is a typical value for the reactor flux normalization uncertainty.

\section{Default energy-scale and flux-shape errors: Hierarchy tests}

In this section we study the hierarchy sensitivity of the reference JUNO experiment, under the effects 
of increasingly large sets of systematic errors. Default $1\sigma$ errors are assumed
for energy-scale and flux-shape deviations (see Fig.~1). The various
effects are first shown graphically and then quantified in
terms of the variable \cite{Stat} $N_\sigma =\sqrt{\Delta \chi^2}$.

Figure~2 (upper panel) shows the absolute NH and IH event spectra $S^\mathrm {NH}$ and $S^\mathrm{IH}$, respectively, as obtained by simply 
swapping the hierarchy, without any change in the central value of the
oscillation parameters or other systematics (case of ``no uncertainties''). For definiteness, the spectra 
correspond to 5 years of data taking. The lower panel of Fig.~2 shows the IH/NH spectral ratio, with characteristic wiggles due
to the mismatch between the oscillation peak positions in the two hierarchies. 
The amplitude of the wiggles reaches $\sim 4\%$ for $E_\mathrm{vis}\sim 2 $--3~MeV. In this ideal situation,
the NH and IH spectra would be distinguishable ``by eye.''

In Fig.~3, the NH spectrum is taken as input data ($S^*=S^\mathrm{NH}$), while the IH spectrum is fitted ($S=S^\mathrm{IH}$), 
with allowance for  oscillation parameter uncertainties and normalization systematics. With respect to Fig.~2,
the mismatch between NH and IH spectra
appears to be reduced to $\lesssim 3\%$ in the low-energy range,  although it slightly increases (up to $\sim 1.5\%$) at high
energy.

Figure~4 is similar to Fig.~3, but with energy-scale uncertainties included in the fit. In this case, the NH and IH spectra are barely
distinguishable by eye, their relative mismatch being lower than $\sim 2\%$ at any energy. This trend is even more pronounced in
Fig.~5, where flux-shape uncertainties have been included in the fit: the NH and IH spectra appear to be almost indistinguishable,
except for percent-level differences in the oscillation peaks around $2$~MeV.

In Figs.~2--5, a sharp rise of the event spectra is evident at the IBD threshold. The steep derivative 
guarantees that energy-scale and flux-shape deviations cannot be large at threshold, otherwise the spectral
mismatch would locally ``explode,'' with a significant $\chi^2$ increase.  In this sense,
the IBD threshold acts as a ``self-calibrating point'': the true (NH) spectrum and the test (IH) spectrum must almost coincide
at such kinematical threshold, and systematic errors in the fit cannot alter this requirement. [At most, 
the small residual mismatch around threshold may lead to locally fuzzy wiggles in the spectral ratio $S^*/S$, as partly 
shown in Figs.~4 and 5.]
   
The above expectations are confirmed by an analysis
of the best-fit energy profiles (fifth-degree polynomials) for the energy-scale and flux-shape deviations,
corresponding to the cases shown in Figs.~3, 4 and 5. Figure~6 shows such profiles (solid curves) superimposed to the default error
bands (in color) for $E'/E$ (top panels) and $\Phi'/\Phi$ (bottom panels). 
The leftmost panels of Fig.~6 correspond to the fit in Fig.~3, which includes only
oscillation and normalization errors. In this case, $E'/E=1$ by construction (no energy-scale error), while $\Phi'/\Phi=1+\beta_0$ 
can float (to account for the flux normalization), but happens to have a best-fit value very close to unity.

\begin{figure}[t]
\hfil
\begin{minipage}[t]{0.46\textwidth}
\includegraphics[width=\textwidth]{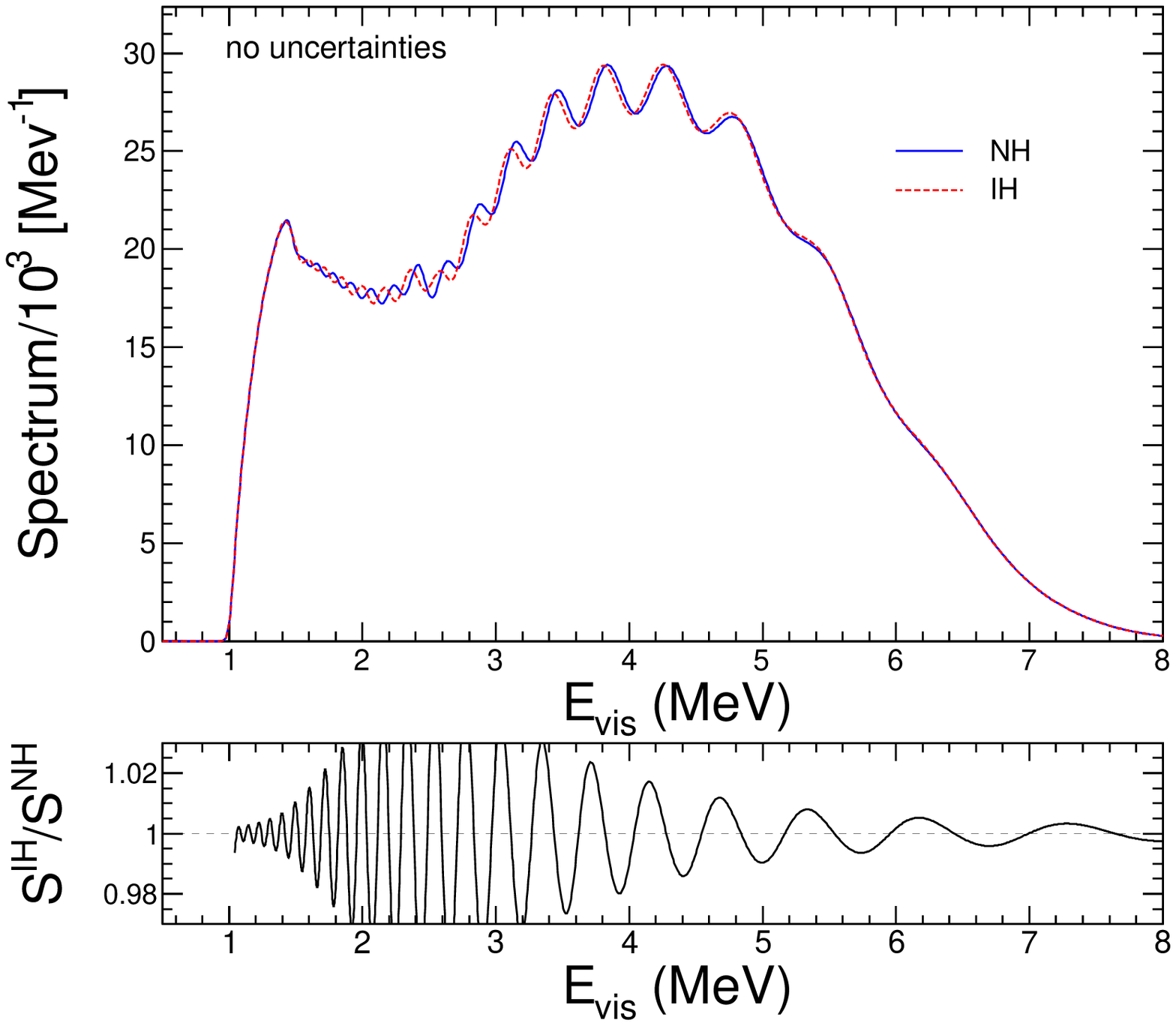}
\caption{\label{fig02}\footnotesize Comparison of event spectra in NH and IH, as obtained for fixed 
oscillation parameters and no systematic errors. Top: Absolute spectra for $T=5$~yr. Bottom: Spectral ratio.}
\end{minipage}
\hspace{0.03\textwidth}
\begin{minipage}[t]{0.46\textwidth}
\includegraphics[width=\textwidth]{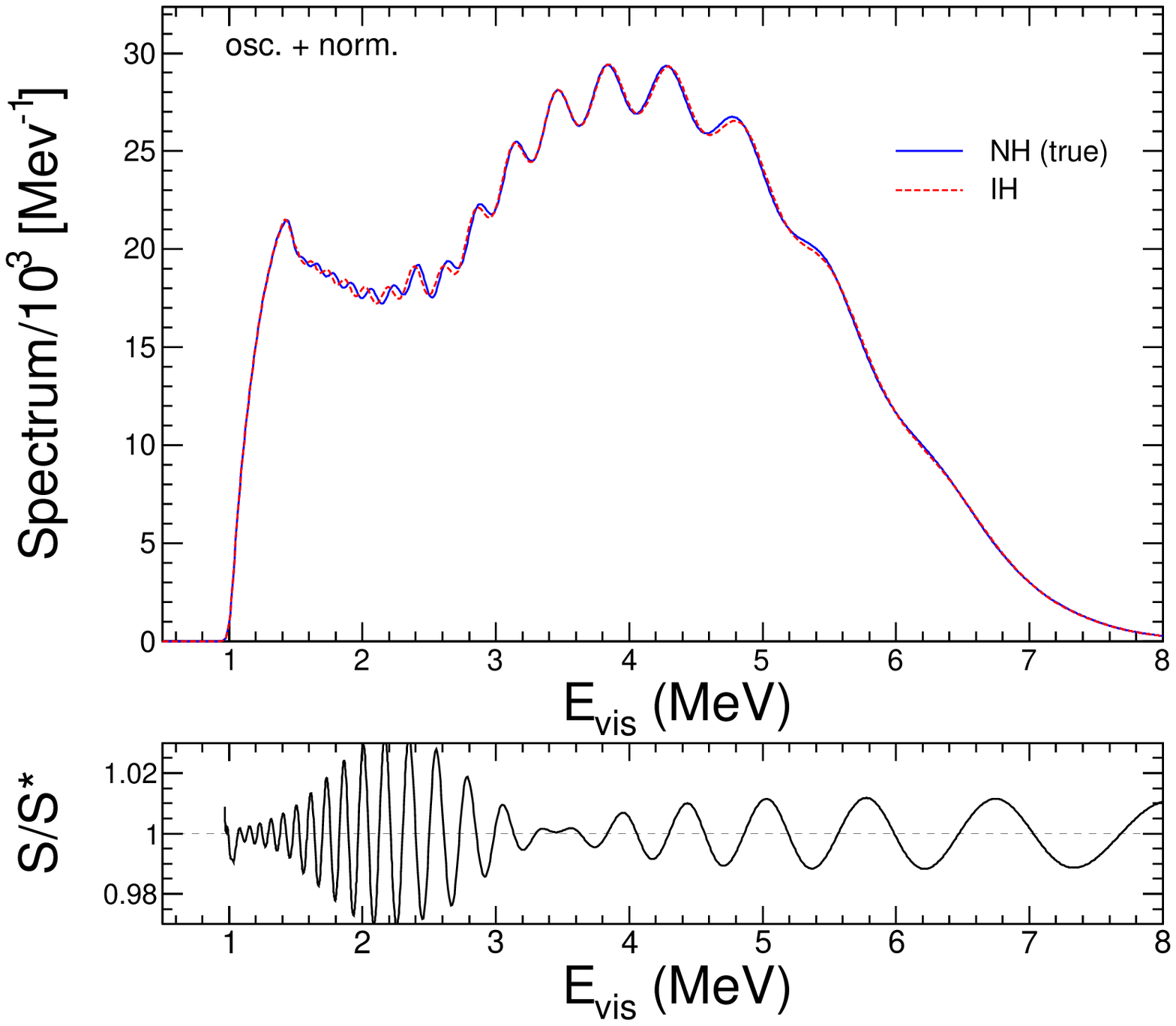}
\caption{\label{fig03}\footnotesize Comparison of event spectra in NH (true, $S^*$) and IH (fitted, $S$),
including oscillation and normalization uncertainties.}
\end{minipage}
\hfil
\end{figure}

\begin{figure}[t]
\hfil
\begin{minipage}[t]{0.46\textwidth}
\includegraphics[width=\textwidth]{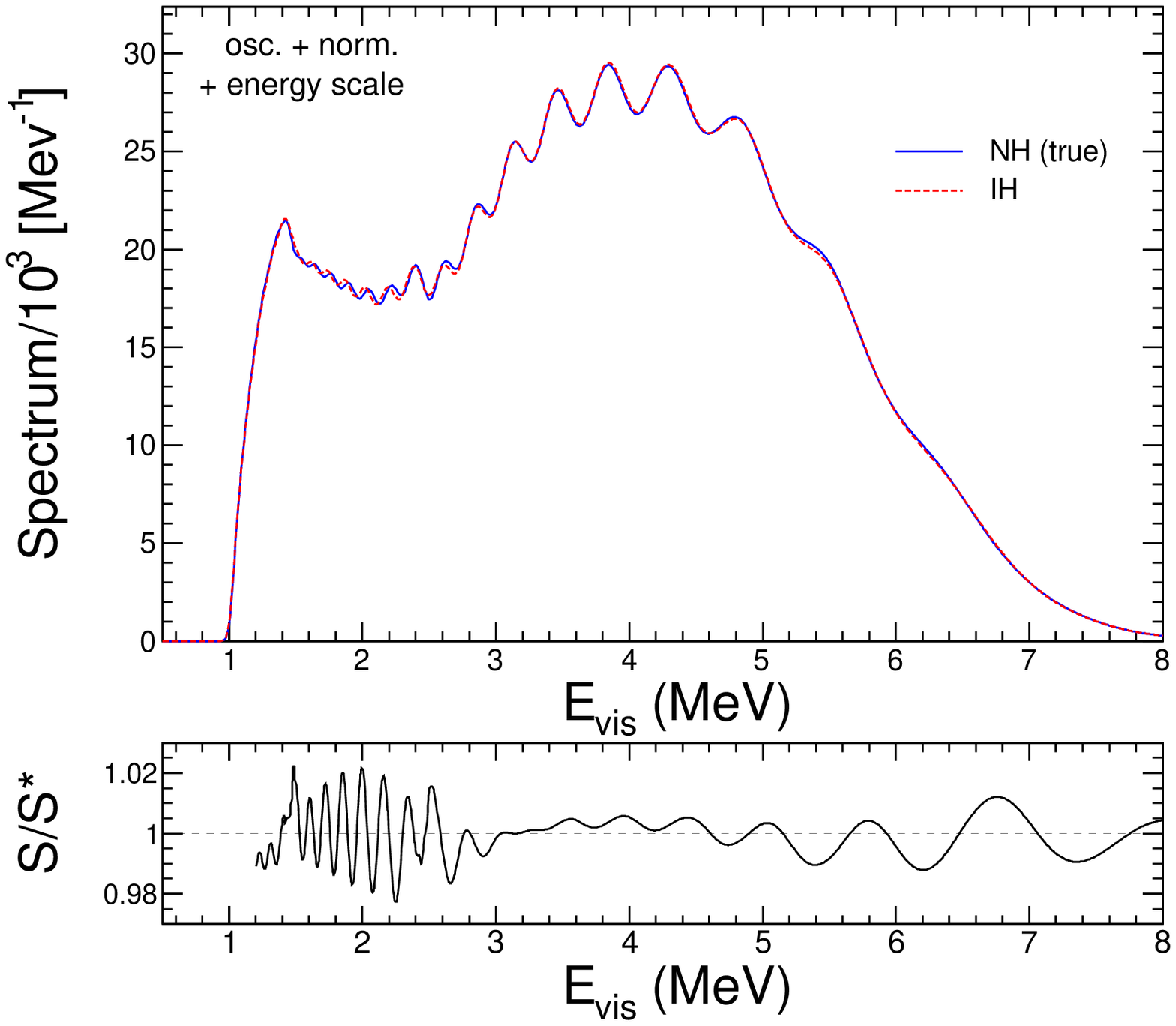}
\caption{\label{fig04}\footnotesize As in Fig.~3, but including energy-scale systematics.}
\end{minipage}
\hspace{0.03\textwidth}
\begin{minipage}[t]{0.46\textwidth}
\includegraphics[width=\textwidth]{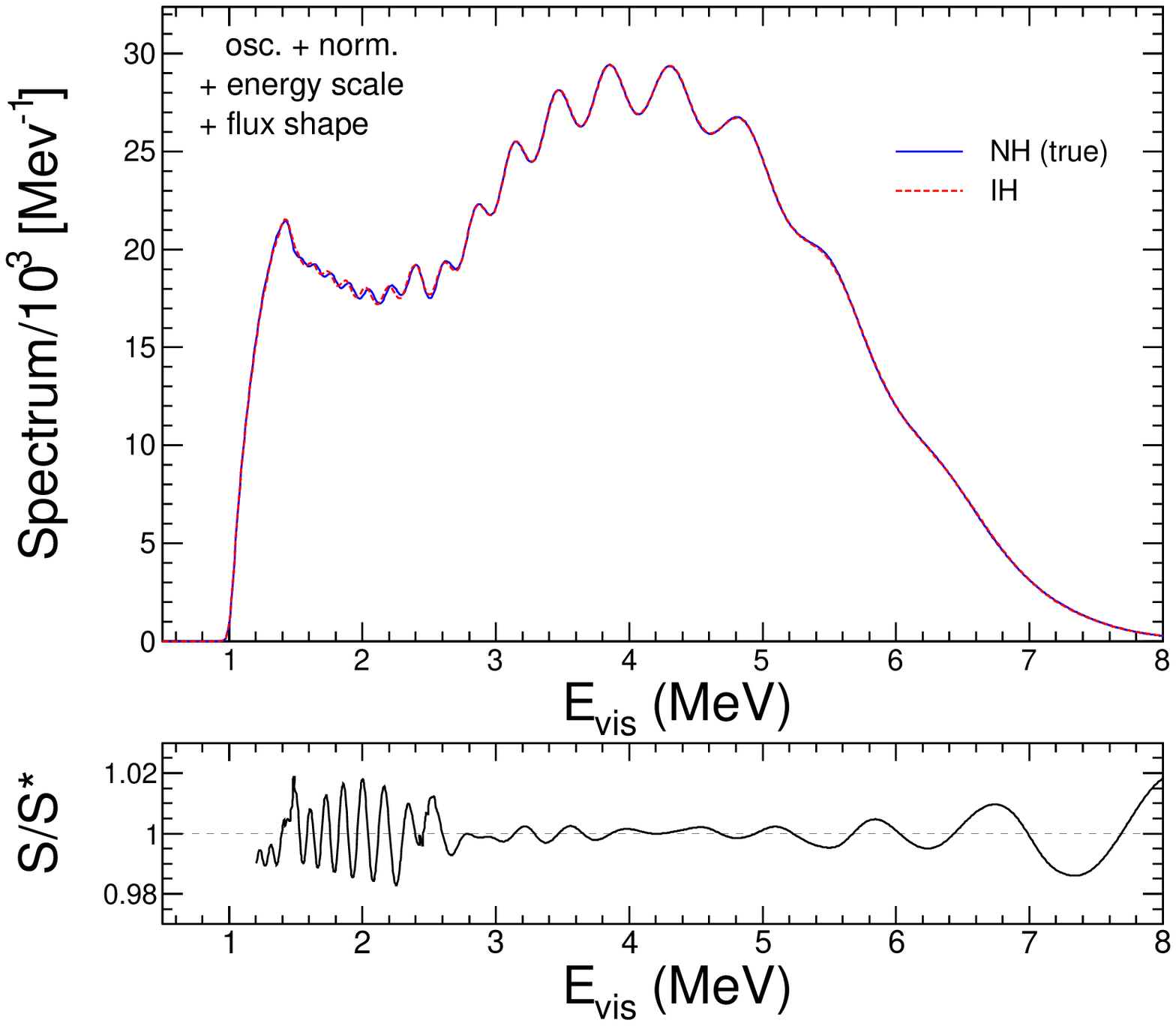}
\caption{\label{fig05}\footnotesize As in Fig.~4, but including flux-shape systematics.}
\end{minipage}
\hfil
\end{figure}

The middle panels in 
Fig.~6 correspond to the fit in Fig.~4, which also includes energy-scale systematics.  In this case, one can observe a slight
offset of the overall ratio $\Phi'/\Phi=1+\beta_0$, and a peculiar pattern for the best-fit $E'/E$ profile.
The function $E'/E$ is close to unity at the IBD threshold ($E\simeq 1.8$~MeV, equivalent to $E_\mathrm{vis}\simeq 1$ MeV), 
according to expectations. Then the function rises up by $\sim +0.6\%$, changes sign at $E\simeq 4$~MeV, decreases
by $\sim -0.6\%$, and approaches unity at high energy. The sign-flip 
behavior is vaguely reminiscent of the ``engineered'' $E'/E$ profile that
would realign, by construction, the IH and NH oscillation phases 
and peaks \cite{Qian} while remaining unity at the IBD threshold (see Fig.~17 in
\cite{Ours}). However, the ``fitted''  $E'/E$ profile in Fig.~6 does not need to reach large deviations at high $E$
as the engineered one \cite{Ours}, since the high-energy tail of the  spectrum 
contributes marginally to the $\chi^2$ function. Finally, the rightmost panels in Fig.~6 correspond to the
complete fit in Fig.~5, which includes also flux-shape systematics. The function $E'/E$ is qualitatively similar
to the middle panel, but with reduced deviations in the high-energy part of the spectrum. The best-fit function $\Phi'/\Phi$
shows sign-changing deviations at the few-percent level, well within the $\pm1\sigma$ (colored) error band. In conclusion,
admissible systematic deformations of the energy scale and of the flux shape, added to the usual
oscillation parameter and normalization uncertainties, may bring the ``true'' and ``wrong'' event spectra
as close to each other as shown in  Fig.~5.

\begin{figure}[t]
\begin{minipage}[c]{0.96\textwidth}
\includegraphics[width=0.54\textwidth]{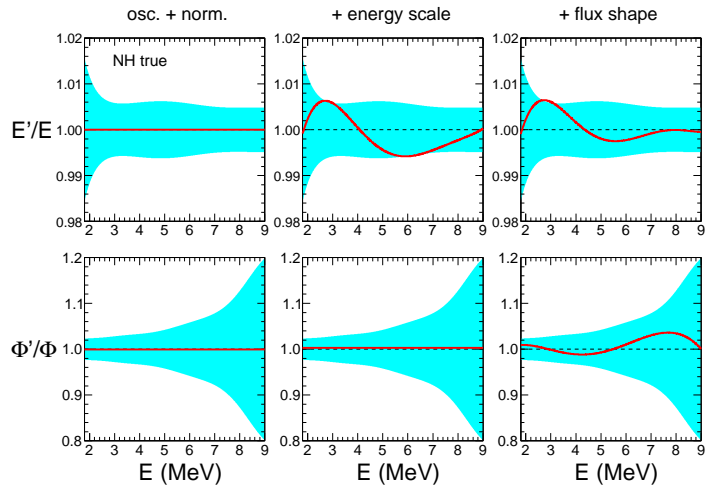}
\caption{\label{fig06}\footnotesize Energy profile of best-fit deviations $E'/E$ (top panels) and $\Phi'/\Phi$ (bottom panels), 
for different sets of systematic uncertainties.}
\end{minipage}
\end{figure}

Figure~7 shows the statistical significance of the wrong hierarchy rejection, for the case of true normal hierarchy, in terms of
$N_\sigma = \sqrt{\Delta\chi^2}$ as a function of live time $T$. The abscissa scales as $\sqrt{T}$,
thus showing at a glance any deviation from the ideal ``linear'' case of purely statistical errors ($N_\sigma\propto \sqrt{T}$).  In the fit
including only oscillation parameter and normalization uncertainties, $N_\sigma$ grows steadily and almost linearly in $\sqrt{T}$ 
along ten years of data taking. However, the inclusion of energy-scale uncertainties provides some bending of the linear rise,
with a noticeable but not dramatic decrease of the statistical significance. In particular, it appears that a $3\sigma$ rejection
is achievable after about six years of data taking. We agree with \cite{Unam} that energy-scale uncertainties, by themselves,  do not represent
a showstopper for JUNO-like experiments. However, Fig.~7 shows that the combination of energy-scale and flux-shape systematics
can be quite sizable: the solid curve for $N_\sigma$ grows much more slowly than $\sqrt{T}$, and remains below 
$3\sigma$ even after ten years
of data taking. Figure~8 shows a very similar behavior, but assuming the IH as true. 
Figures~7 and 8 represent 
one of the main results of our work, as they clearly 
demonstrate the importance of
accounting for nonlinear deformations of the spectra $\Phi(E)$ both in abscissa ($E\to E'$) 
and in ordinate ($\Phi \to \Phi'$).

\begin{figure}[b]
\hfil
\begin{minipage}[t]{0.46\textwidth}
\includegraphics[width=\textwidth]{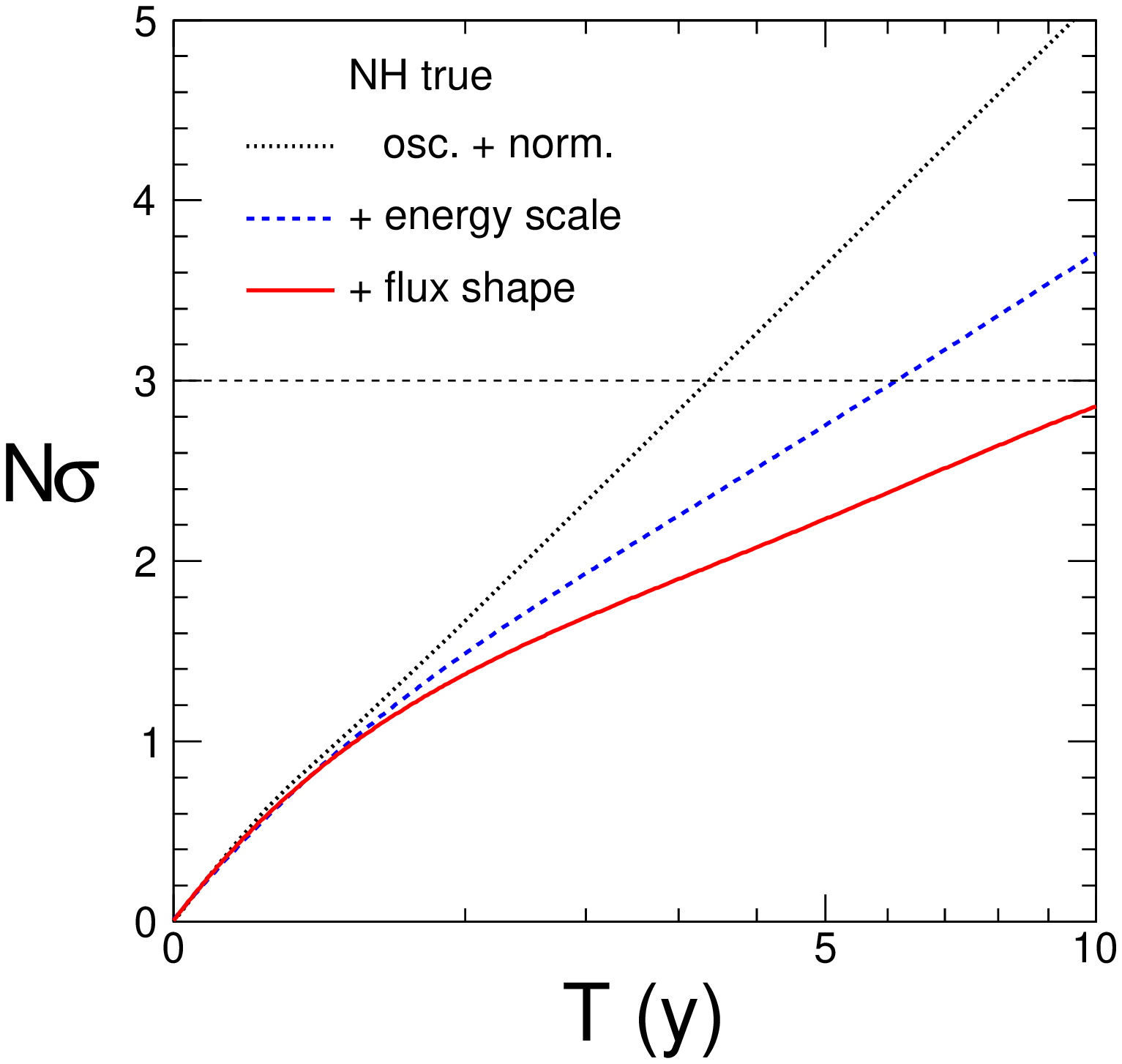}
\caption{\label{fig07}\footnotesize Case of true NH: Statistical significance of the IH
rejection as a function of the detector live time $T$, as derived from fits including 
different sets of systematics. Note that the abscissa scales as $\sqrt{T}$. 
The horizontal $3\sigma$ line is shown to guide the eye.}
\end{minipage}
\hspace{0.03\textwidth}
\begin{minipage}[t]{0.46\textwidth}
\includegraphics[width=\textwidth]{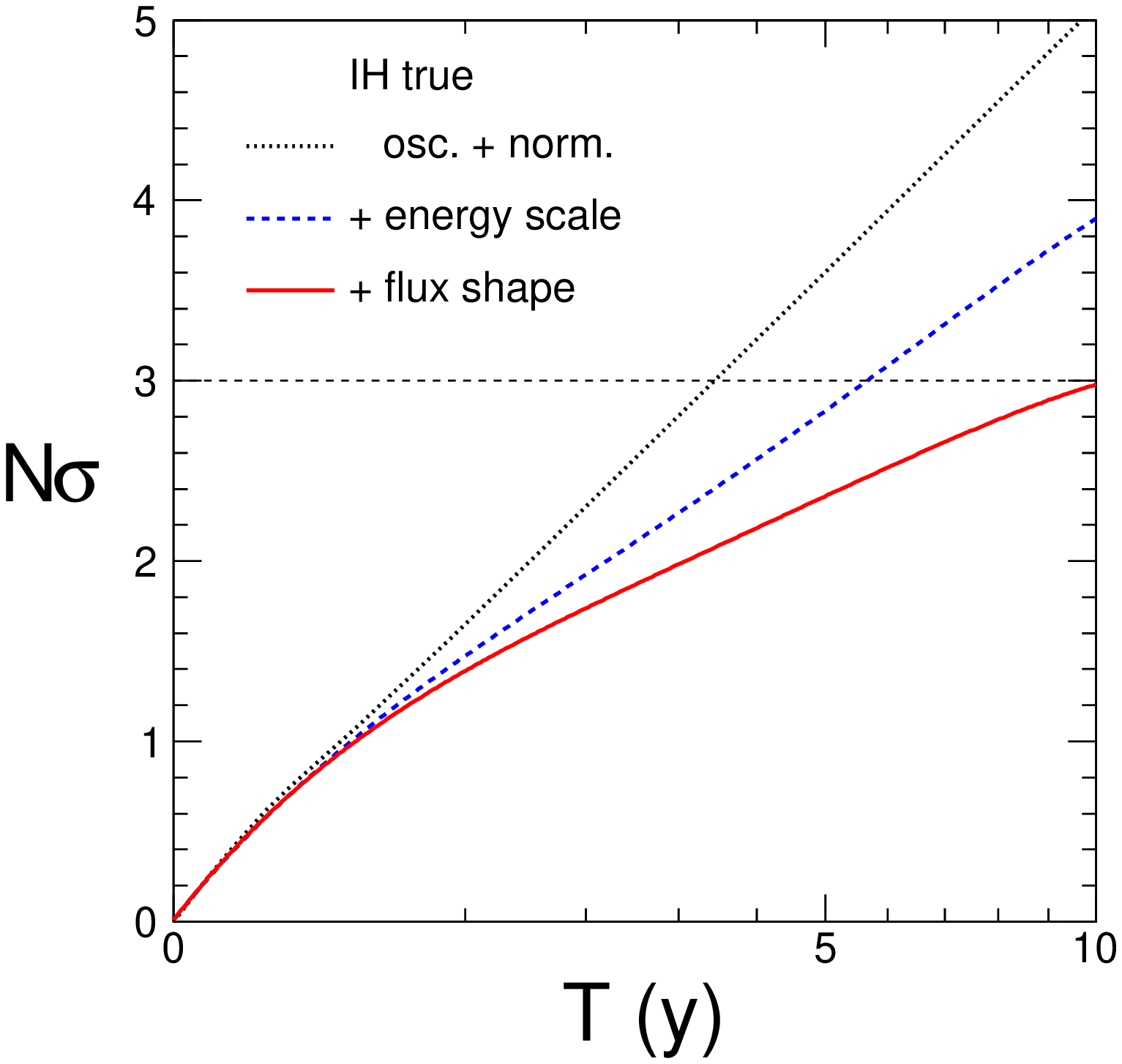}
\caption{\label{fig08}\footnotesize As in Fig.~7, but for true IH and rejection of NH.}
\end{minipage}
\hfil
\end{figure}

\section{Default energy-scale and flux-shape errors: Precision physics}

\begin{figure}[t]
\hfil
\begin{minipage}[t]{0.46\textwidth}
\includegraphics[width=\textwidth]{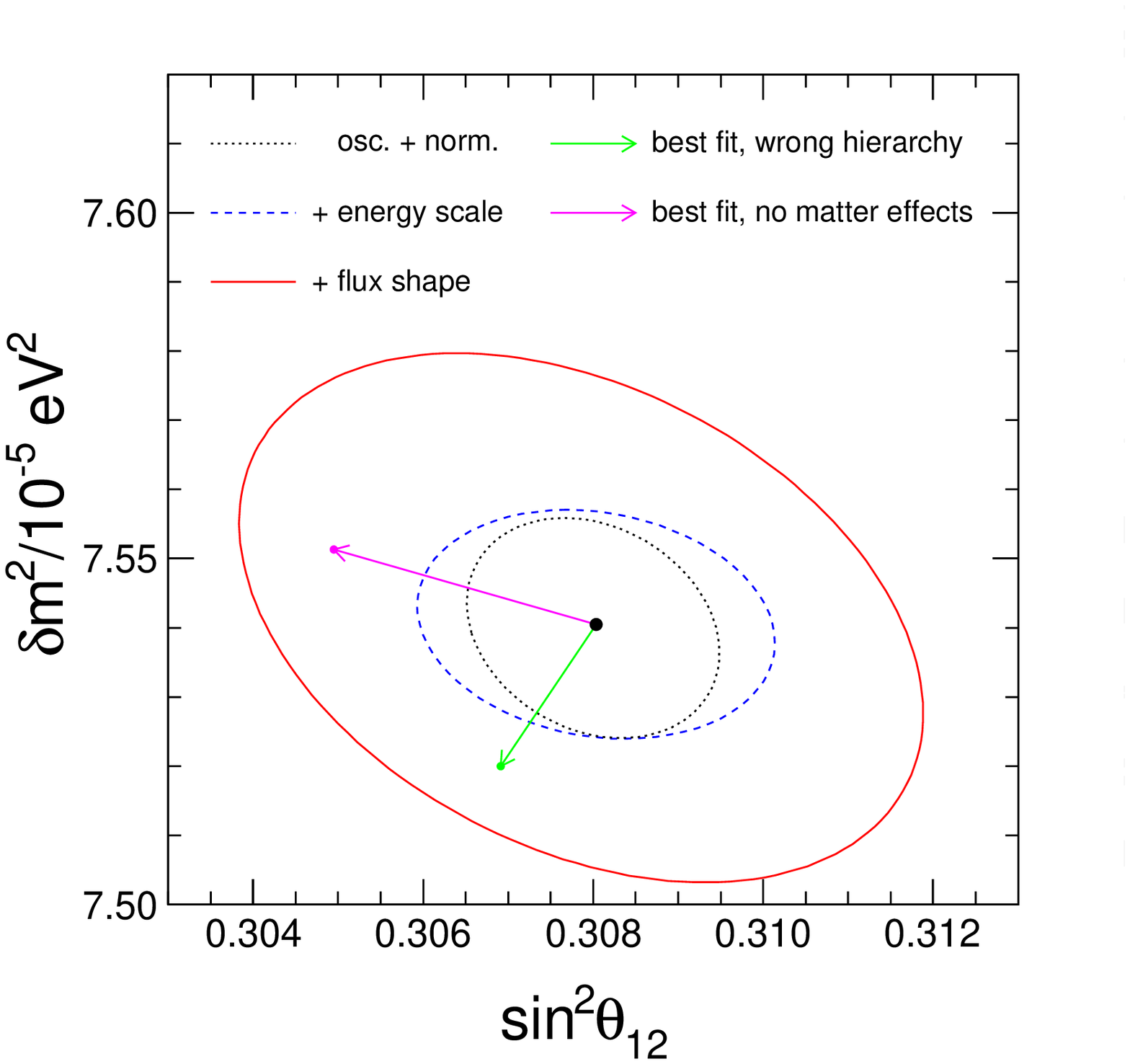}
\caption{\label{fig09}\footnotesize Mass-mixing parameters $(\delta m^2,\,s^2_{12})$: $1\sigma$ contours
for true NH and  $T=5$~y, as derived from fits including 
different systematic uncertainties. The arrows indicate the best-fit displacement in the cases of wrong hierarchy (green)
and no matter effects (magenta).} 
\end{minipage}
\hspace{0.03\textwidth}
\begin{minipage}[t]{0.46\textwidth}
\includegraphics[width=\textwidth]{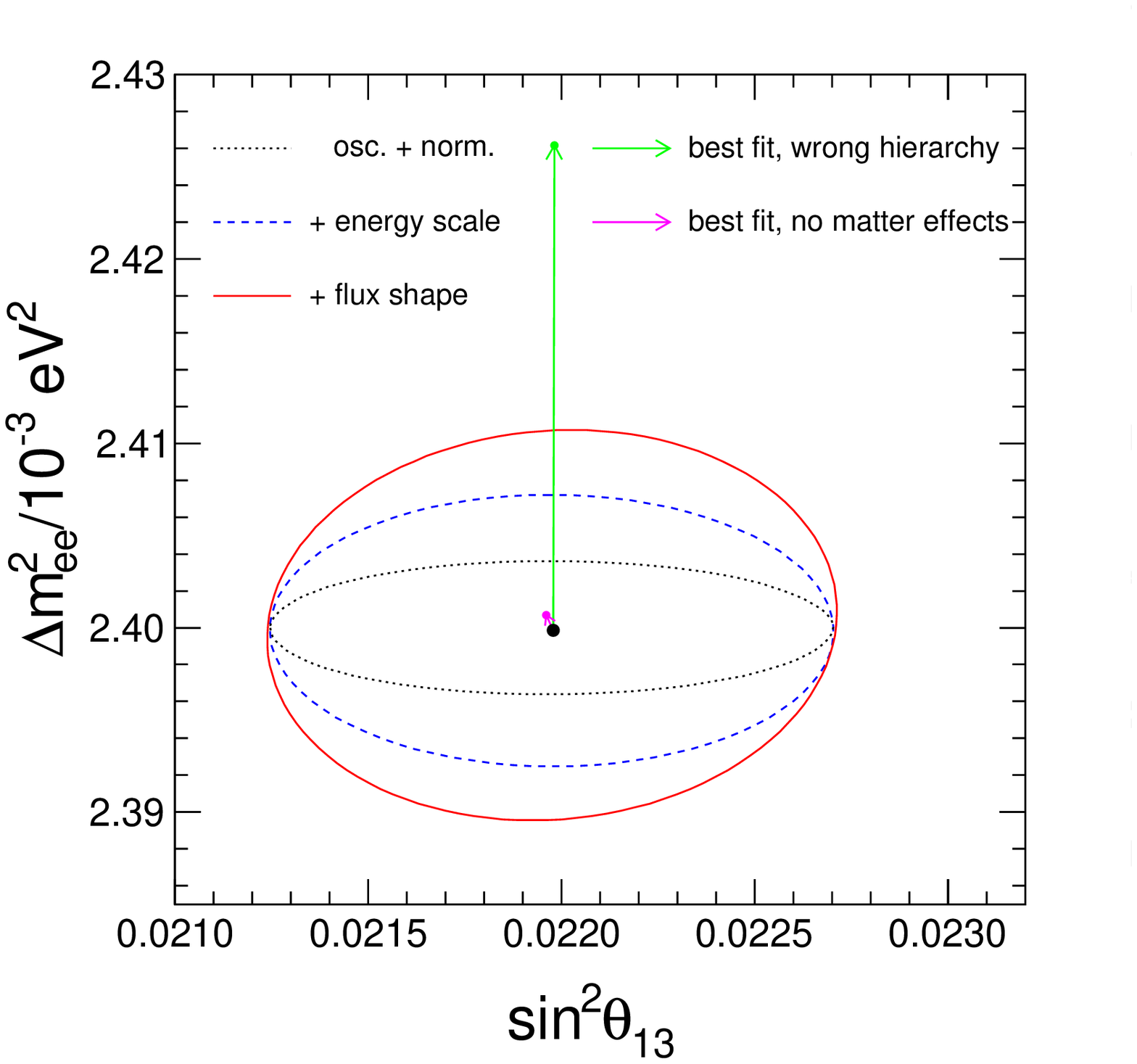}
\caption{\label{fig10}\footnotesize As in Fig.~9, but for the $(\Delta m^2_{ee},\,s^2_{13})$ parameters.}
\end{minipage}
\hfil 
\end{figure}

In this section, unless otherwise noticed, the true hierarchy is assumed to be known, and the discussion is focused on the precision physics program, namely, on the accuracy expected at $\pm1\sigma$ on the oscillation parameters and geoneutrino normalizations.  For definiteness, the accumulated statistics refers
to $T=5$~yr.

Figure~9 shows the $1\sigma$ contours in the plane charted by the mass-mixing parameters $(\delta m^2,\,s^2_{12})$, 
assuming true NH.
In the case with only oscillation and normalization errors, the accuracy
of both $\delta m^2$ and $s^2_{12}$ is more than an order of magnitude better than the prior errors assumed in Sec.~II~A.
The accuracy is slightly worse in the presence of energy-scale systematics, but is significantly degraded 
(by almost a factor of three) in combination with flux-shape systematics. This is not surprising, 
since the $(\delta m^2,\,s^2_{12})$ parameters
govern the long-wavelength oscillation pattern, which is sensitive to smooth deformations  of the spectral shape.  
Note that, in all the above cases of Fig.~9, the best-fit coordinates coincide---by construction---with the 
central values of $(\delta m^2,\,s^2_{12})$ assumed as priors. However, the best-fit point would be significantly 
displaced if the inverse hierarchy were mistakenly assumed as ``true,'' as indicated by the green arrow in Fig.~9 (color online).
A different but still sizable displacement (indicated by the magenta arrow) would also be induced 
by discarding matter effects  in the fit. For simplicity, the matter density has been assumed herein to be constant \cite{Ours} but, in the future,
one should characterize more precisely the density profile from the
geophysical and geochemical viewpoint.
 Summarizing, Fig.~9 shows that  the $1\sigma$ accuracy of the $(\delta m^2,\,s^2_{12})$
measurements in a JUNO-like experiment can be significantly degraded by the combined effect of energy-scale and flux-shape
uncertainties, and that their central values may be biased by
ambiguities in the hierarchy, as well as by ``vacuum'' approximations in the oscillation probability.

Figure~10 shows the $1\sigma$ contours in the plane charted by the mass-mixing parameters $(\Delta m^2_{ee},\,s^2_{13})$, 
assuming true NH. The accuracy on $s^2_{13}$ is essentially constant and almost equal to the prior assignment 
in Sec.~II~A, implying that a JUNO-like experiment cannot really improve the input $\theta_{13}$ data from current  short-baseline reactor experiment. In the case with only oscillation and normalization errors, the accuracy
of $\Delta m^2_{ee}$ is more than an order of magnitude better than the prior error assumed in Sec.~II~A. 
However, the accuracy is degraded (by a factor of two) by energy-scale uncertainties and (by a factor of three) by adding flux-shape
systematics. In fact these systematics, as shown in Sec.~III, may slightly alter the pattern of short-wavelength oscillations and may thus
affect the measurement of its peak frequency, governed by $\Delta m^2_{ee}$. 
Moreover, the central value of $\Delta m^2_{ee}$ (but not of $\theta_{13}$) would be strongly biased in the case of ``wrong'' hierarchy (green arrow), roughly by $(c^2_{12}-s^2_{12})\delta m^2$ as expected from Eq.~(\ref{Deltam2ee}). The vacuum approximation bias (magenta arrow) is instead insignificant,
since the parameters $(\Delta m^2_{ee},\,s^2_{13})$
are basically unaffected by matter effects  \cite{Ours}.
Results similar to Figs.~9 and 10 also hold in the case of true IH (not shown).

\begin{figure}[t]
\begin{minipage}[c]{0.96\textwidth}
\includegraphics[width=0.42\textwidth]{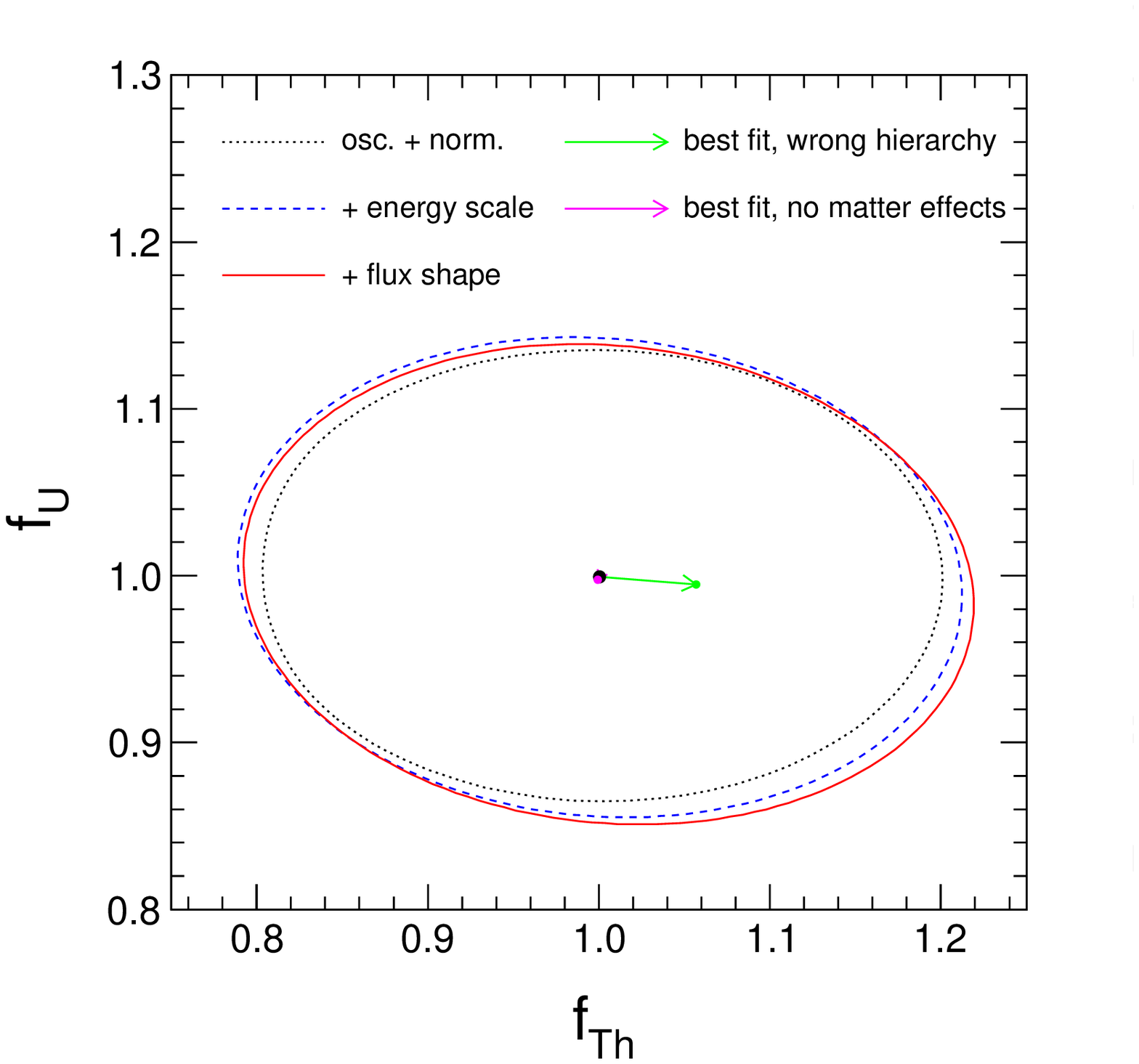}
\caption{\label{fig11}\footnotesize As in Fig.~9, but for (Th,~U)  geo-$\nu$ normalizations.}
\end{minipage}
\end{figure}

Finally, Fig.~11 shows the $1\sigma$ contours in the plane charted by the Th and U geoneutrino flux normalizations
$(f_\mathrm{Th},\,f_\mathrm{U})$. The fit constrains these normalizations within $1\sigma$ errors which are 
smaller (by about $30\%$) than their prior values as defined in Sec.~II~D, and are quite insensitive  to different sources
of systematic errors and biases. Therefore, prospective geoneutrino 
results in JUNO will help to constrain better the current geophysical and geochemical models for the radiogenic element abundances,
independently of systematic details.

\begin{figure}[b]
\begin{minipage}[c]{0.96\textwidth}
\includegraphics[width=0.46\textwidth]{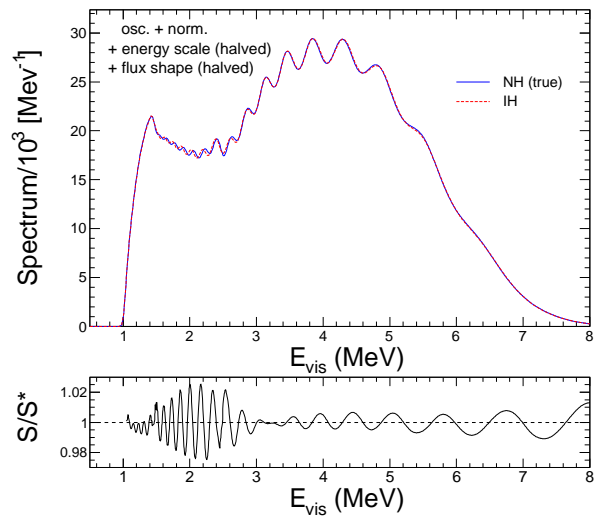}
\caption{\label{fig12}\footnotesize As in Fig.~5, but with halved energy-scale and flux-shape errors.}
\end{minipage}
\end{figure}

\section{Results for halved energy-scale and flux-shape errors}

The reference error bands in Fig.~1 are representative of present state-of-the-art systematic uncertainties on energy-scale and flux-shape deformations
of typical reactor event spectra. When one or more JUNO-like experiments will be operative, it is conceivable that the detector energy scale will be
subject to dedicated calibration campaigns, and that the reactors spectral profiles will be better understood, on the basis of the high-statistics data sets
collected by current-generation short-baseline experiments. Therefore, it makes sense to repeat the analyses in Secs.~III and IV in
the hypothesis of smaller (for definiteness, halved) error bands in Fig.~1, while all other priors are assumed to be unchanged. The results are discussed
below.

\begin{figure}[t]
\begin{minipage}[c]{0.96\textwidth}
\includegraphics[width=0.54\textwidth]{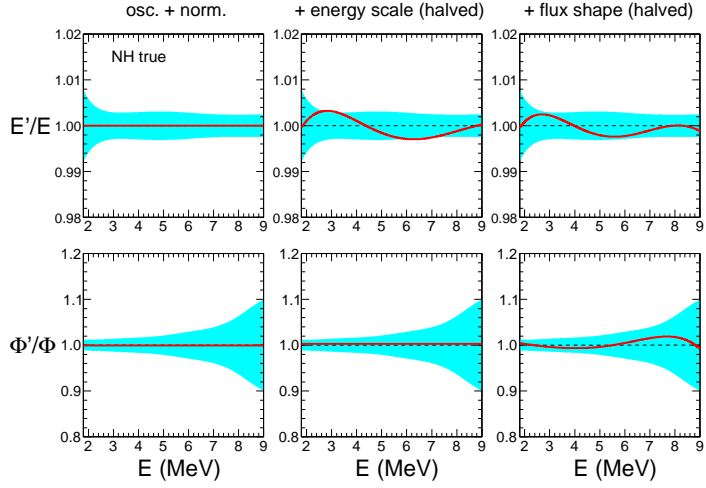}
\caption{\label{fig13}\footnotesize As in Fig.~6, but with halved energy-scale and flux-shape uncertainties.}
\end{minipage}
\vspace*{-2mm}
\end{figure}

\subsection{Hierarchy tests}
\vspace*{-2mm}

Figure~12 is analogous to Fig.~5, but with halved energy-scale and flux-shape uncertainties. It can be seen that 
the NH and IH spectra cannot be brought as close to each other as in Fig.~5, and that the residual spectral differences are noticeably larger,
as a result of the systematic error reduction. 

Figure~13 shows the effect of halving errors on the energy profiles of the energy-scale and flux-shape deformations,
corresponding to the best-fit IH spectrum in Fig.~12. From the comparison of Fig.~6 and Fig.~13, it appears that the qualitative 
behavior of the profiles is similar for either default or halved errors but, in the latter case, the amplitude is suppressed
by roughly a factor of two.

\begin{figure}[b]
\hfil
\begin{minipage}[t]{0.46\textwidth}
\includegraphics[width=\textwidth]{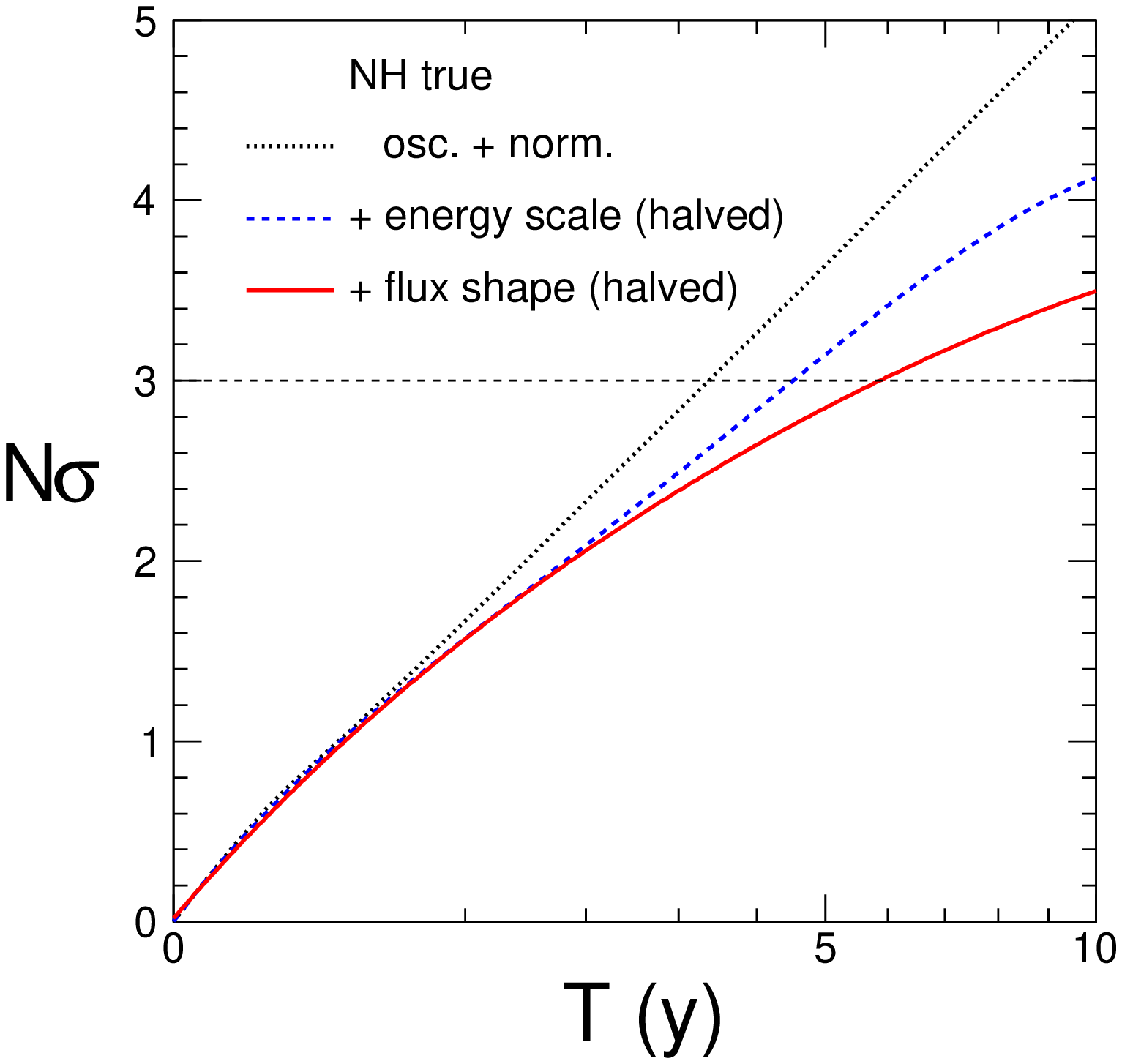}
\caption{\label{fig14}\footnotesize As in Fig.~7 (true NH), but with halved energy-scale and flux-shape uncertainties.}
\end{minipage}
\hspace{0.03\textwidth}
\begin{minipage}[t]{0.46\textwidth}
\includegraphics[width=\textwidth]{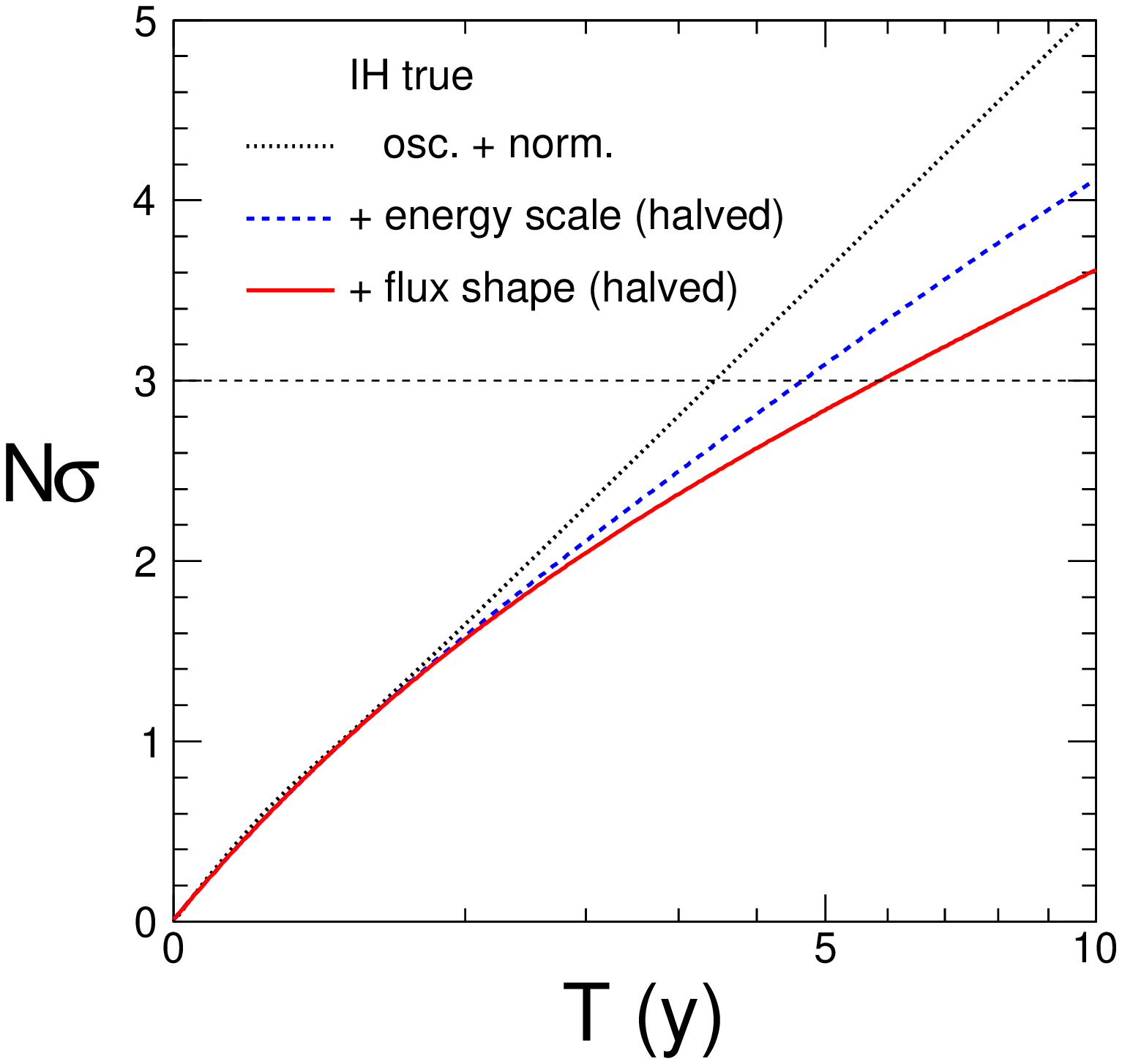}
\caption{\label{fig15}\footnotesize As in Fig.~14, but for true IH.}
\end{minipage}
\hfil
\end{figure}

From Figs.~12 and 13, one expects a non-negligible improvement in testing the wrong hierarchy versus the true one. Indeed, Figs.~14 and 15 
show the statistical significance of the wrong-hierarchy rejection, in the cases with true NH and true IH, respectively: in both cases,
a $3\sigma$ rejection level appears to be reachable in about 6 years of data taking, consistently with the expected goal of a JUNO-like
experiment \cite{Unam}. By comparing Figs.~14 and 15 with the 
analogous Figs.~7 and 8, one can derive the following conclusions: (1) energy-scale and flux-shape uncertainties
tend to decrease by comparable amounts the statistical significance of the hierarchy test; (2) such errors, 
according to current estimates, may prevent (in combination) an effective hierarchy discrimination; (3) a 
future error reduction by a factor of two may lead to a $\gtrsim 3\sigma$ rejection of the wrong hierarchy, with a
reasonable detector exposure ($T\simeq 6$~y); (4) a significance $\gtrsim 4\sigma$ seems to be out of reach for a one-decade
exposure, unless all systematics 
are further reduced.

\begin{figure}[t]
\hfil
\begin{minipage}[t]{0.46\textwidth}
\includegraphics[width=\textwidth]{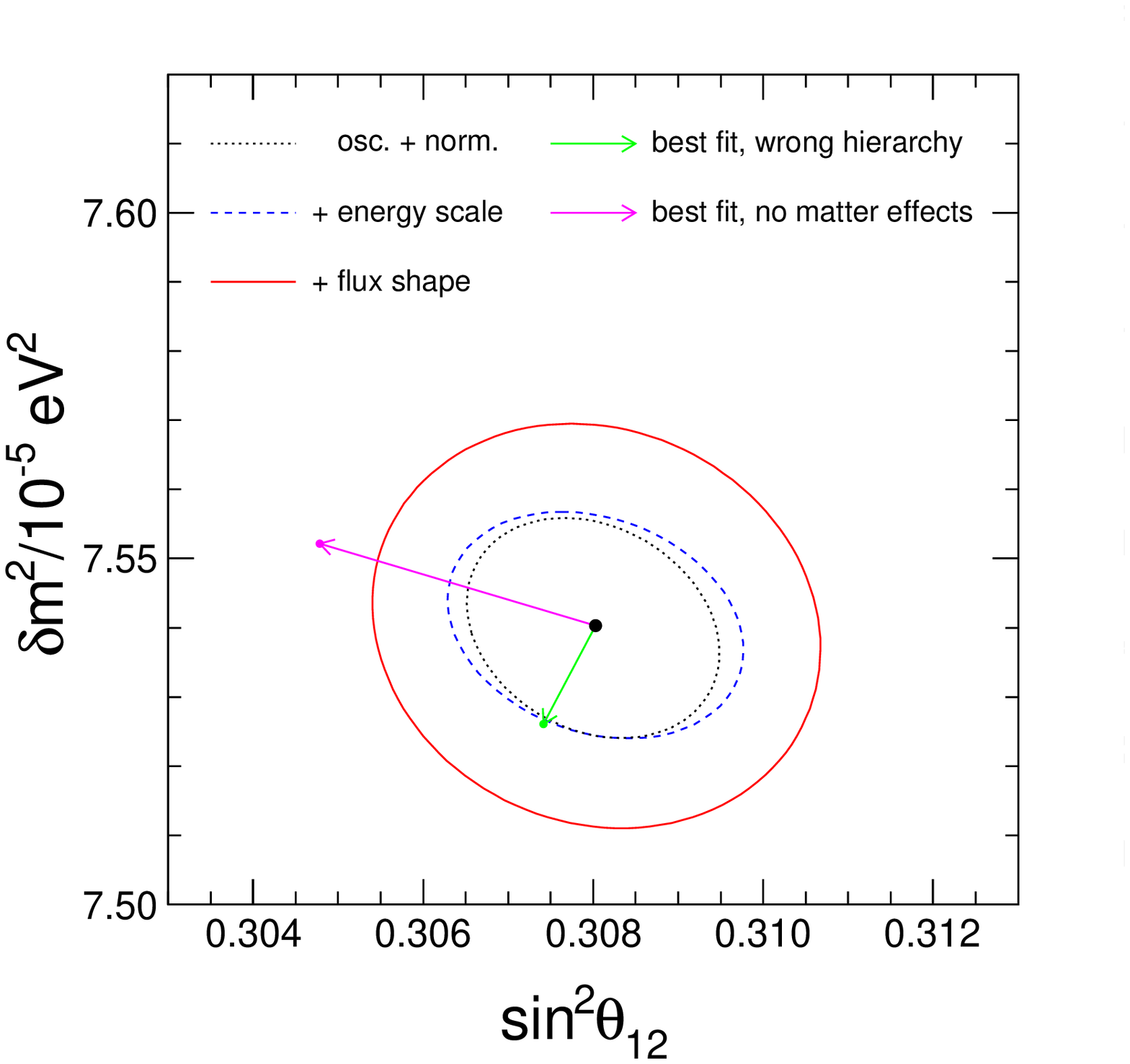}
\caption{\label{fig16}\footnotesize As in Fig.~9, but with halved energy-scale and flux-shape uncertainties.}
\end{minipage}
\hspace{0.03\textwidth}
\begin{minipage}[t]{0.46\textwidth}
\includegraphics[width=\textwidth]{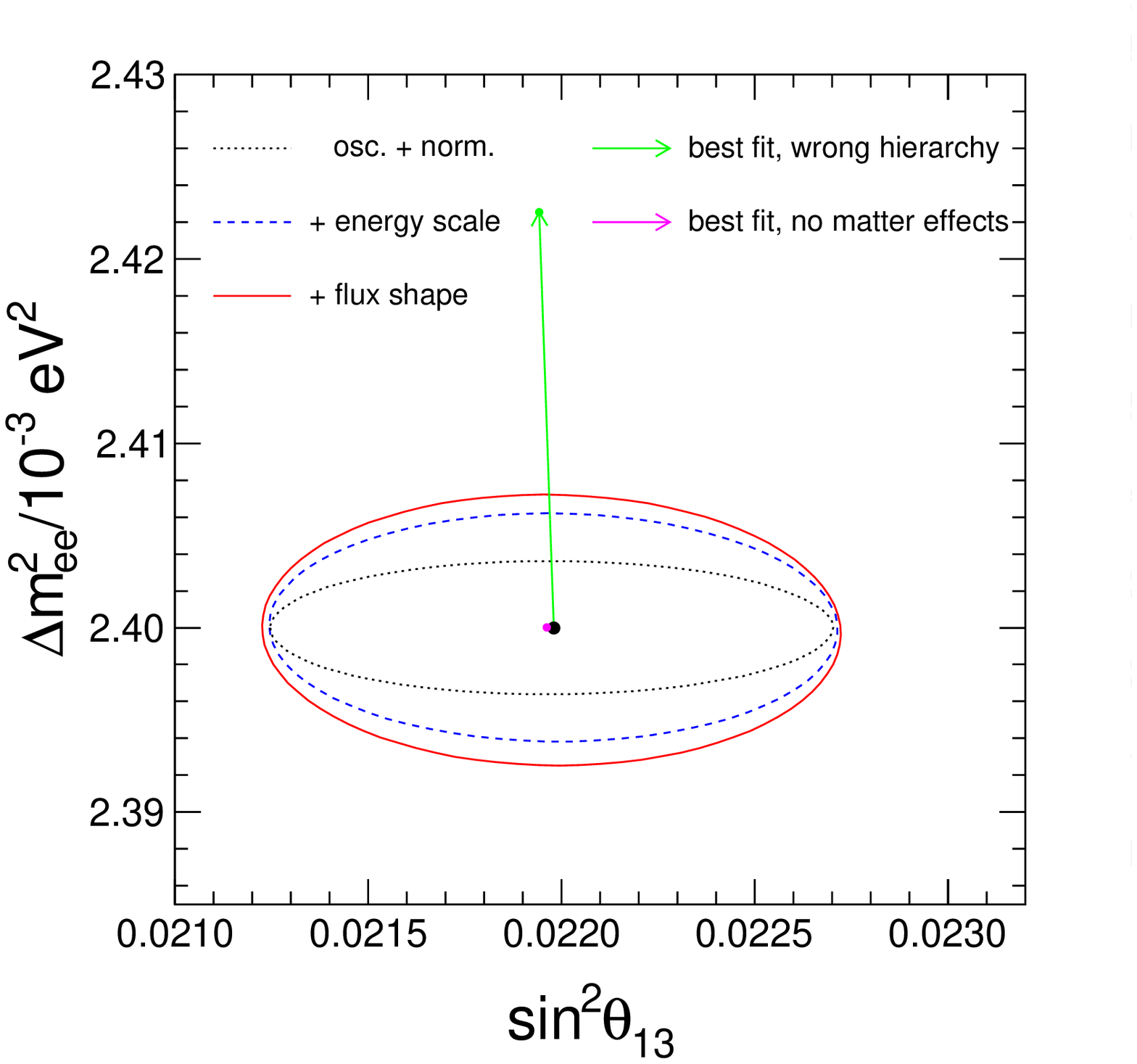}
\caption{\label{fig17}\footnotesize As in Fig.~10, but with halved energy-scale and flux-shape uncertainties.}
\end{minipage}
\hfil
\end{figure}

\vspace*{-5mm}
\subsection{Precision physics}
\vspace*{-1mm}

Halving the energy-scale and flux-shape uncertainties has also a significant impact on the precision program in a JUNO-like experiment.
Figures~16 and 17 (to be compared with the analogous Figs.~9 and 10, respectively), show that such uncertainties, 
with respect to the case with only oscillation and normalization errors, degrade the final accuracy on the $(\delta m^2,\,s^2_{12},\,\Delta m^2_{ee})$
parameters by a factor of two or less, while the $s^2_{13}$ accuracy remains always close to its prior assignment. 
The constraints on the Th and U geoneutrino flux normalizations (not shown) are basically the same as in Fig.~11, since
these spectral components are quite insensitive to details of the energy scale and flux shape.

Table~I summarizes the fit results for the oscillation and the geoneutrino parameters, in terms of (symmetrized) $1\sigma$ errors, to be compared with 
their prior  $\pm 1\sigma$ ranges. The results refer to NH and to increasing sets of systematic uncertainties, including 
energy-scale and flux-shape errors at default values (see Figs.~9 and 10) and halved values (see Figs.~16 and 17). 
In conclusion, 
reducing the error bands in Fig.~1 represents a major requirement to fully exploit the physics potential of a JUNO-like
experiment, both for discriminating the neutrino mass hierarchy and for measuring more precisely the $(\delta m^2,\,s^2_{12},\,\Delta m^2_{ee})$
oscillation parameters and the geoneutrino fluxes.
\vspace*{-2mm}

\begingroup
\squeezetable
\begin{table}[bh]
\caption{\scriptsize Precision physics in a JUNO-like experiment, assuming known normal hierarchy.
1st and 2nd column: oscillation or geoneutrino parameter, together with the assumed prior value and $\pm 1\sigma$ error.
3rd column: $1\sigma$ error from the fit to prospective 5-year data, including only oscillation and normalization uncertainties. 4th and 5th column:
$1\sigma$ error from the fit, including also energy-scale and flux-shape uncertainties with default error bands. 6th and 7th column: as in the previous two columns, but with halved error bands. Similar results are obtained for the case of known inverted hierarchy 
 (not shown). See the text for details.}
\centering 
\begin{ruledtabular}
\begin{tabular}{ccccccc}
Parameter & Prior $\pm 1\sigma$  & Osc.~+~norm. & +~Energy scale & +~Flux shape & +~Energy scale & +~Flux shape \\
		  &                      &  fit error   &  (default)     & (default) & (halved) & (halved) \\
\hline 
$s^2_{12}/10^{-1}$ 						& $3.08\pm 0.17$ & 0.015  & 0.021  & 0.040 & 0.017  & 0.026   \\
$\delta m^2/10^{-5}\mathrm{\ eV}^2$ 	& $7.54\pm 0.20$ & 0.016  & 0.017  & 0.038 & 0.016  & 0.029   \\
$s^2_{13}/10^{-2}$ 						& $2.20\pm 0.08$ & 0.073  & 0.073  & 0.074 & 0.074  & 0.074   \\
$\Delta m^2_{ee}/10^{-3}\mathrm{\ eV}^2$& $2.40\pm 0.05$ & 0.0036 & 0.0074 & 0.011 & 0.0064 & 0.0074  \\
$f_\mathrm{Th}$ 						& $1.00\pm 0.27$ & 0.20   & 0.21   & 0.21  & 0.21   & 0.21    \\
$f_\mathrm{U}$ 							& $1.00\pm 0.20$ & 0.14   & 0.14   & 0.14  & 0.14   & 0.14 
\label{tab1}
\end{tabular}
\end{ruledtabular}
\end{table} 
\endgroup

\newpage

\section{Summary and conclusions}

Medium baseline, high-statistics reactor neutrino projects such as JUNO (in construction) and RENO-50 (proposed) 
can pursue an important research program in neutrino physics, including the
determination of the unknown mass hierarchy, precision measurements of some known oscillation parameters, 
and improved constraints on geoneutrino fluxes.
In this context,  building upon our previous work \cite{Ours},
we have examined in detail the effects of nonlinear variations of the energy scale, $E\to E'(E)$, and
of the unoscillated reactor neutrino flux shape, $\Phi(E)\to \Phi'(E)$, in addition to the usual prior uncertainties associated with oscillation and 
normalization parameters. For definiteness, we have performed our analysis in a JUNO-like configuration,  
assuming  energy-scale and flux-shape error bands anchored to
state-of-the-art estimates (default case), as well as for error bands reduced by a factor of two  (halved case), as shown in Fig.~1.
 
It turns out that such systematics can noticeably affect the performance of the experiment, and that their reduction is
mandatory in order to achieve statistically significant results, both for hierarchy discrimination and for precision physics. 
In particular, a $>3\sigma$ separation of NH and IH might not be reached after one decade in the case of
default systematic errors (Figs.~7 and 8),
while it can be reached after $\sim 6$ years of data taking in the case of halved errors (Figs.~14 and 15).
Similarly, assuming that the hierarchy is known, the energy-scale and flux-shape systematic uncertainties 
can significantly affect the accuracy of the ($s^2_{12},\,\delta m^2,\,\Delta m^2_{ee}$) oscillation parameters 
emerging from prospective data fits (see Table~I). 

The main message of our work is that further constraints on the admissible shapes and sizes of $E\to E'(E)$ 
and $\Phi(E)\to \Phi'(E)$ variations  would be highly beneficial to the entire physics program of medium-baseline reactor projects. 
As side results of our analysis, we find the following: (1) the well-known IBD energy threshold acts as an effective self-calibration point in the fit; (2) neglecting matter effects may significantly bias the oscillation parameters ($s^2_{12},\,\delta m^2$); (3) taking the wrong hierarchy may significantly bias the parameter $\Delta m^2_{ee}$; and (4) prospective constraints on Th and U geoneutrino fluxes are largely insensitive to systematic uncertainties.


\acknowledgments

This work is supported by the Italian Istituto Nazionale di Fisica 
Nucleare (INFN) and Ministero dell'Istruzione, Universit\`a e Ricerca (MIUR) through the ``Theoretical Astroparticle Physics''  projects.
Preliminary results  have been presented by A.M.\ at {\em EPS-HEP 2015\/}, European Physical Society Conference on High Energy Physics
(Vienna, Austria, July 2015).



\begin{thebibliography}{99}

\bibitem{PDGR} K.A. Olive et al. (Particle Data Group), Chin.\ Phys.\ C {\bf 38}, 090001 (2014). See the review therein:  
``Neutrino mass, mixing and oscillations,'' by  K.\ Nakamura and S.T.\ Petcov.


\bibitem{Fo01} 
  G.~L.~Fogli, E.~Lisi and A.~Palazzo,
  ``Quasi energy independent solar neutrino transitions,''
  Phys.\ Rev.\ D {\bf 65}, 073019 (2002)
  [hep-ph/0105080].

\bibitem{Pe02} 
  S.~T.~Petcov and M.~Piai,
  ``The LMA MSW solution of the solar neutrino problem, inverted neutrino mass hierarchy and reactor neutrino experiments,''
  Phys.\ Lett.\ B {\bf 533}, 94 (2002)
  [hep-ph/0112074].


\bibitem{JUN1} 
  F.~An {\it et al.},
  ``Neutrino Physics with JUNO,''
  arXiv:1507.05613 [physics.ins-det].

\bibitem{RE50}
  S.~B.~Kim,
  ``New results from RENO and prospects with RENO-50,''
  arXiv:1412.2199 [hep-ex], to appear in the Proceedings of
  {\em NOW 2014\/}, Neutrino Oscillation Workshop (Otranto, Italy, 2014), ed.
  by P.~Bernardini, G.L.~Fogli and E.~Lisi, Nucl.\ Part.\ Phys.\ Proc.\ 
  (Elsevier, 2015, in press).

\bibitem{Gosw}  A.~Bandyopadhyay, S.~Choubey and S.~Goswami,
 ``Exploring the sensitivity of current and future experiments to theta(solar),''
 Phys.\ Rev.\ D {\bf 67}, 113011 (2003)
 [hep-ph/0302243];   
%
 S.~Choubey, S.~T.~Petcov and M.~Piai,
  ``Precision neutrino oscillation physics with an intermediate baseline reactor neutrino experiment,''
  Phys.\ Rev.\ D {\bf 68}, 113006 (2003)
  [hep-ph/0306017];
%
 A.~Bandyopadhyay, S.~Choubey, S.~Goswami and S.~T.~Petcov,
 ``High precision measurements of theta(solar) in solar and reactor neutrino experiments,''
 Phys.\ Rev.\ D {\bf 72} (2005) 033013
 [hep-ph/0410283].


\bibitem{Min1}
  H.~Minakata, H.~Nunokawa, W.~J.~C.~Teves and R.~Zukanovich Funchal,
  ``Reactor measurement of theta(12): Principles, accuracies and physics potentials,''
  Phys.\ Rev.\ D {\bf 71}, 013005 (2005)
  [hep-ph/0407326].


\bibitem{Vo15} 
  P.~Vogel, L.~Wen and C.~Zhang,
  ``Neutrino Oscillation Studies with Reactors,''
  Nature Communications {\bf 6}, 6935
  (2015) [arXiv:1503.01059 [hep-ex]].

\bibitem{Vo16}
  X.~Qian and P.~Vogel,
  ``Neutrino Mass Hierarchy,''
  Prog.\ Part.\ Nucl.\ Phys.\  {\bf 83}, 1 (2015)
  [arXiv:1505.01891 [hep-ex]].


\bibitem{BRen}
  S.~H.~Seo [RENO Collaboration],
  ``New Results from RENO and The 5 MeV Excess,''
  Proceedings of {\em Neutrino 2014\/}, XXVI International Conference
  on Neutrino Physics and Astrophysics, ed by E.\ Kearns and G.\ Feldman,
  AIP Conf.\ Proc.\  {\bf 1666}, 080002 (2015)
  [arXiv:1410.7987 [hep-ex]].


\bibitem{BDou}  Y.~Abe {\it et al.} [Double Chooz Collaboration],
  ``Improved measurements of the neutrino mixing angle $\theta_{13}$ with the Double Chooz detector,''
  JHEP {\bf 1410}, 086 (2014)
  [Erratum ibidem {\bf 1502}, 074 (2015)]
  [arXiv:1406.7763 [hep-ex]].


\bibitem{BDay}
  L.~Zhan [for the Daya Bay Collaboration],
  ``Recent Results from Daya Bay,''
  arXiv:1506.01149 [hep-ex], to appear in the Proceedings of 
  {\em NEUTEL 2015\/}, XVI International  Workshop on Neutrino Telescopes
  (Venice, Italy, 2015).
  

\bibitem{Muel}
  T.~A.~Mueller {\it et al.},
  ``Improved Predictions of Reactor Antineutrino Spectra,''
  Phys.\ Rev.\ C {\bf 83}, 054615 (2011)
  [arXiv:1101.2663 [hep-ex]].

\bibitem{Hube}
  P.~Huber,
  ``On the determination of anti-neutrino spectra from nuclear reactors,''
  Phys.\ Rev.\ C {\bf 84}, 024617 (2011)
  [Erratum ibidem {\bf 85}, 029901 (2012)]
  [arXiv:1106.0687 [hep-ph]].

\bibitem{Dwye}
  D.~A.~Dwyer and T.~J.~Langford,
  ``Spectral Structure of Electron Antineutrinos from Nuclear Reactors,''
  Phys.\ Rev.\ Lett.\  {\bf 114}, no. 1, 012502 (2015)
  [arXiv:1407.1281 [nucl-ex]].


\bibitem{Haye}
  A.~C.~Hayes, J.~L.~Friar, G.~T.~Garvey, D.~Ibeling, G.~Jungman, T.~Kawano and R.~W.~Mills,
  ``The Origin and Implications of the Shoulder in Reactor Neutrino Spectra,''
  arXiv:1506.00583 [nucl-th].

\bibitem{HuGe} P.~Huber, talk at the International Conference {\em Neutrino Geoscience 2015\/} (Paris, France, 2015), available at the website
 www.ipgp.jussieu.fr/fr/evenements/neutrino-geoscience-2015-conference
 


\bibitem{DB15}
  F.~P.~An {\it et al.} [Daya Bay Collaboration],
  ``A new measurement of antineutrino oscillation with the full detector configuration at Daya Bay,''
  arXiv:1505.03456 [hep-ex].

\bibitem{Park}
  S.~J.~Parke, H.~Minakata, H.~Nunokawa and R.~Z.~Funchal,
  ``Mass Hierarchy via Mossbauer and Reactor Neutrinos,''
  Nucl.\ Phys.\ Proc.\ Suppl.\  {\bf 188}, 115 (2009)
  [arXiv:0812.1879 [hep-ph]].


\bibitem{Qian} 
  X.~Qian, D.~A.~Dwyer, R.~D.~McKeown, P.~Vogel, W.~Wang and C.~Zhang,
  ``Mass Hierarchy Resolution in Reactor Anti-neutrino Experiments: Parameter Degeneracies and Detector Energy Response,''
  Phys.\ Rev.\ D {\bf 87}, no. 3, 033005 (2013)
  [arXiv:1208.1551 [physics.ins-det]].

\bibitem{Unam} 
  Y.~F.~Li, J.~Cao, Y.~Wang and L.~Zhan,
  ``Unambiguous Determination of the Neutrino Mass Hierarchy Using Reactor Neutrinos,''
  Phys.\ Rev.\ D {\bf 88}, 013008 (2013)
  [arXiv:1303.6733 [hep-ex]].



\bibitem{Ours} 
  F.~Capozzi, E.~Lisi and A.~Marrone,
  ``Neutrino mass hierarchy and electron neutrino oscillation parameters with one hundred thousand reactor events,''
  Phys.\ Rev.\ D {\bf 89}, no. 1, 013001 (2014)
  [arXiv:1309.1638 [hep-ph]].

\bibitem{PING} 
  F.~Capozzi, E.~Lisi and A.~Marrone,
  ``PINGU and the neutrino mass hierarchy: Statistical and systematic aspects,''
  Phys.\ Rev.\ D {\bf 91}, 073011 (2015)
  [arXiv:1503.01999 [hep-ph]].

\bibitem{Gouv}
  A.~de Gouvea, J.~Jenkins and B.~Kayser,
  ``Neutrino mass hierarchy, vacuum oscillations, and vanishing |U(e3)|,''
  Phys.\ Rev.\ D {\bf 71}, 113009 (2005)
  [hep-ph/0503079].

\bibitem{Nuno}
   H.~Nunokawa, S.~J.~Parke and R.~Zukanovich Funchal,
  ``Another possible way to determine the neutrino mass hierarchy,''
  Phys.\ Rev.\ D {\bf 72}, 013009 (2005)
  [hep-ph/0503283].

\bibitem{Mina}
  H.~Minakata, H.~Nunokawa, S.~J.~Parke and R.~Zukanovich Funchal,
  ``Determination of the neutrino mass hierarchy via the phase of the disappearance oscillation probability with a monochromatic anti-electron-neutrino source,''
  Phys.\ Rev.\ D {\bf 76}, 053004 (2007)
  [Erratum-ibid.\ D {\bf 76}, 079901 (2007)]
  [hep-ph/0701151].


\bibitem{GFit}  F.~Capozzi, G.~L.~Fogli, E.~Lisi, A.~Marrone, D.~Montanino and A.~Palazzo,
  ``Status of three-neutrino oscillation parameters, circa 2013,''
  Phys.\ Rev.\ D {\bf 89}, 093018 (2014)
  [arXiv:1312.2878 [hep-ph]].


\bibitem{Wang} Y.\ Wang, talk at the International Conference on Massive Neutrinos (Nanyang Technological Univ., Singapore, 2015), available
at the website: www.ntu.edu.sg/ias/upcomingevents/MassiveNeutrinos

\bibitem{Fit1}  D.~V.~Forero, M.~Tortola and J.~W.~F.~Valle,
  ``Neutrino oscillations refitted,''
  Phys.\ Rev.\ D {\bf 90}, no. 9, 093006 (2014)
  [arXiv:1405.7540 [hep-ph]].


\bibitem{Fit2}   M.~C.~Gonzalez-Garcia, M.~Maltoni and T.~Schwetz,
  ``Updated fit to three neutrino mixing: status of leptonic CP violation,''
  JHEP {\bf 1411}, 052 (2014)
  [arXiv:1409.5439 [hep-ph]].


\bibitem{T2K1}
  K.~Abe {\it et al.} [T2K Collaboration],
  ``Neutrino oscillation physics potential of the T2K experiment,''
  Prog.\ Theor.\ Exp.\ Phys.\ {\bf 2015}, no.~4, 043C01 (2015)
  [arXiv:1409.7469 [hep-ex]].

\bibitem{JGeo}
V.~Strati, M.~Baldoncini, I.~Callegari, F.~Mantovani, W.F.~McDonough, B.~Ricci and
G.~Xhixha, Progress in Earth and Planetary Science {\bf 2}, article ID~5, (2015) [arXiv:1412.3324 [physics.geo-ph]].

\bibitem{Err1} D.A.~Dwyer, talk at the Workshop ``The Status of Reactor Antineutrino Flux Modeling'' (Nantes, France, 2014),
available at indico.cern.ch/event/353976

\bibitem{Err2} B.-Z.\ Hu, talk at {\em Moriond EW 2015\/}, 50th Rencontres de Moriond on ElectroWeak Interactions and Unified Theories (La Thuile, Italy, 2015),
available at indico.in2p3.fr/event/10819

\bibitem{Ge12} 
  S.~F.~Ge, K.~Hagiwara, N.~Okamura and Y.~Takaesu,
  ``Determination of mass hierarchy with medium baseline reactor neutrino experiments,''
  JHEP {\bf 1305}, 131 (2013)
  [arXiv:1210.8141 [hep-ph]].

\bibitem{Expl} In our opinion, it is appropriate to attach flux-shape uncertainties to the ``theoretical spectrum'' $S$ and energy-scale uncertainties to the
``experimental spectrum'' $S^*$. However, we have verified that our results are basically unchanged, if both uncertainties are assumed to act only on $S$ or on $S^*$. In such cases, in principle, one must also specify the ordering of the non-commutative operations $E\to E'$ and $\Phi \to \Phi'$. We have also verified that commuting such operations (on either $S$ or $S^*$) induces negligible numerical changes in our results.  

\bibitem{Stat} 
  M.~Blennow, P.~Coloma, P.~Huber and T.~Schwetz,
  ``Quantifying the sensitivity of oscillation experiments to the neutrino mass ordering,''
  JHEP {\bf 1403}, 028 (2014)
  [arXiv:1311.1822 [hep-ph]].
 

\end{thebibliography}
\end{document}